\documentclass[showpacs,aps,prd,preprint,nofootinbib,showkeys,unsortedaddress,raggedbottom]{revtex4-1}
\pdfoutput=1

\usepackage{lipsum}

\usepackage{bm}
\usepackage{amsmath}
\usepackage{amssymb}
\usepackage{graphicx}
\usepackage{subfigure}
\usepackage[usenames,dvipsnames]{color}
\definecolor{darkblue}{RGB}{0,0,196}
\usepackage[colorlinks=true,linkcolor=darkblue,citecolor=darkblue,urlcolor=darkblue]{hyperref}
\usepackage{diagbox}

\usepackage{setspace}
\usepackage{footmisc}
\usepackage[makeroom]{cancel}
  
\DeclareMathOperator{\arctanh}{arctanh}

\def\be{\begin{equation}}
\def\ee{\end{equation}}
\def\ba{\begin{eqnarray}}
\def\ea{\end{eqnarray}}

\newcommand{\vE}{\vec E}
\newcommand{\rh}{\hat r}
\renewcommand{\vr}{\vec r}
\newcommand{\vp}{\vec p}
\newcommand{\eps}{{\varepsilon}}
\newcommand{\rmii}[1]{{\mbox{\tiny\rm{#1}}}}
\newcommand{\mD}{m_\rmii{D}}
\renewcommand{\Re}{\rm Re}
\renewcommand{\Im}{\rm Im}

\begin{document}

\title{Bottomonium suppression using a lattice QCD vetted potential}

\author{Brandon Krouppa} 
\affiliation{Department of Physics, Kent State University, Kent, OH 44242 United States}

\author{Alexander Rothkopf} 
\affiliation{Institute for Theoretical Physics, Heidelberg University, Philosophenweg 12, 69120 Heidelberg, Germany}

\author{Michael Strickland} 
\affiliation{Department of Physics, Kent State University, Kent, OH 44242 United States}

\begin{abstract}
We estimate bottomonium yields in relativistic heavy-ion collisions using a lattice QCD vetted, complex-valued, heavy-quark potential embedded in a realistic, hydrodynamically evolving medium background. We find that the lattice-vetted functional form and temperature dependence of the proper heavy-quark potential dramatically reduces the dependence of the yields on parameters other than the temperature evolution, strengthening the picture of bottomonium as QGP thermometer. Our results also show improved agreement between computed yields and experimental data produced in RHIC 200 GeV/nucleon collisions. For LHC 2.76 TeV/nucleon collisions, the excited states, whose suppression has been used as a vital sign for quark-gluon-plasma production in a heavy-ion collision, are reproduced better than previous perturbatively-motivated potential models; however, at the highest LHC energies our estimates for bottomonium suppression begin to underestimate the data.  Possible paths to remedy this situation are discussed.
\end{abstract}

\date{\today}



\maketitle



\section{Introduction}
\label{sec:intro}

Ultrarelativistic heavy-ion collision (URHIC) experiments being performed at the Large Hadron Collider (LHC) at CERN and the Relativistic Heavy Ion Collider (RHIC) at Brookhaven National Laboratory aim to recreate a primordial state of nuclear matter known as the quark-gluon plasma (QGP). Comparisons between hydrodynamic simulations that describe the evolution of the bulk matter and experimental data suggest that LHC URHICs produce a QGP with an initial temperature on the order of $T_{0}=600{-}700$ MeV at $\tau_0 = 0.25$ fm/c \cite{Romatschke:2009im,Heinz:2013AnnRevNuc,Schenke:2011tv,*Gale:2013IntJModPhys,*Ryu:2015vwa,Niemi:2015qia,*Niemi:2015voa,Alqahtani:2017jwl,Alqahtani:2017tnq}, for beam energies in the range of 2.76 - 5.02 TeV/nucleon.  In addition, analysis of the collective flow of the matter produced in URHICs indicates that the QGP behaves like a nearly inviscid relativistic fluid \cite{Romatschke:2009im,Schenke:2011tv,Heinz:2013AnnRevNuc,Gale:2013IntJModPhys,Ryu:2015vwa,Niemi:2015qia,Niemi:2015voa,Alqahtani:2017jwl,*Alqahtani:2017tnq}.  Although the light hadronic states, such as pions, are disassociated at temperatures around the pseudocritical temperature $T_c \simeq 155$ MeV, it was predicted that, due to their large binding energies, bound states of heavy quarks could survive up to temperatures on the order of 600 MeV.  Due to their short formation times, such heavy-quark bound states probe the history of the QGP and their suppression relative to production in pp collisions was proposed to be a clear signal of the creation of the QGP stemming from temperature-dependent screening of the color force \cite{Matsui:1986dk,Karsch:1987pv}.  In practice, experimentalists do see a reduced yield of heavy quarkonium compared to elementary collisions both at RHIC and LHC \cite{Andronic:2015wma,Mocsy:2013syh}. 

As we have learned over the course of the last years, however, the suppression of heavy quarkonium may be accompanied by the regeneration of the observed quarkonium states (a) at the phase boundary between the QGP and the hadronic phase \cite{BraunMunzinger:2000ep,Thews:2000rj,Grandchamp:2002wp} or (b) dynamically due to in-medium recombination of $Q\bar{Q}$ pairs \cite{Young:2008he,*Young:2009tj,Emerick:2011xu}. If a sizable number of $Q\bar{Q}$ pairs is created in the initial stages of a collision, the probability for a free quark antiquark pair to coalesce into a bound state at some point during the QGP lifetime can become significant. Measurements of charmonium yields e.g. have shown that for this lighter flavor indeed regeneration becomes quite important at LHC energies \cite{Rapp:2017chc}. For bottomonium, only recently have there been studies carried out that include a regeneration component, see e.g. \cite{Du:2017qkv}.  In the case of bottomonium, one expects regeneration to be less important due to its much heavier rest mass and large vacuum binding energy ($\lesssim$ 1 GeV) and this is borne out by detailed calculations.  That said, it is desirable to have a unified framework that includes the effects of both in-medium suppression and regeneration.

Recent measurements at LHC have also shown an unambiguous signal for the elliptic flow of the $J/\Psi$ particle, which implies that charm quarks are at least partially kinetically equilibrated with the surrounding medium \cite{ALICE:2013xna,Liu:2009gx,Zhao:2012gc}. A similar observation for bottomonium has not been made and it is expected that bottomonium does not yet participate in the collective motion of the bulk at 5.02 TeV. Since equilibration is intimately related to a loss of memory, charmonium at LHC appears to provide us with a window into the late stages of the collision, while bottomonium is considered to act as probe of the full evolution of the QGP.  As a result, a detailed understanding of heavy quarkonium in-medium will open the possibility to infer the time-dependent properties of the nuclear matter produced in a heavy-ion collision. It is an open question which properties the different species are most sensitive to, as candidates of course temperature or the shear viscosity to entropy ratio come to mind. 

In this paper, we focus solely on bottomonium and investigate its suppression at both RHIC and LHC energies. To this end we combine a non-relativistic description of the quark-antiquark bound state in terms of an in-medium potential with a realistic dynamical model of the bulk matter created in the collision.   The potential model used is vetted by comparing it to lattice QCD calculations of the real and imaginary parts of the heavy-quark potential, providing the first model to attempt to constrain the full complex potential using lattice input.  In addition, since bottomonium states are believed to be formed early on in a collision ($\tau < 1$ fm/c), they can be sensitive to the early-time non-equilibrium dynamics of the QGP.  Of particular importance is the large pressure anisotropy of the QGP in the local rest frame, ${\cal P}_L \ll {\cal P}_T$, which is induced by the rapid longitudinal expansion of the QGP created in URHICs \cite{Ryblewski:2013jsa,Strickland:2014pga}.  This pressure anisotropy leads to potentially important non-equilibrium corrections to the widths of the various bottomonia states \cite{Dumitru:2007hy,Burnier:2009yu,Dumitru:2009fy,Strickland:2011mw,Strickland:2011aa,Krouppa:2015yoa,Du:2016wdx,Krouppa:2016jcl,Du:2016wdx,Biondini:2017qjh,Krouppa:2017lsw,Nopoush:2017zbu} which we will take into account using information provided by perturbative calculations of the screening masses in an anisotropic QGP \cite{Romatschke:2003ms,Dumitru:2007hy,Dumitru:2009fy,Strickland:2011aa,Nopoush:2017zbu} coupled with a realistic 3+1d dissipative anisotropic hydrodynamic evolution for the QGP background~\cite{Martinez:2010sc,Florkowski:2010cf,Nopoush:2014pfa,Alqahtani:2015qja,Alqahtani:2016rth,Alqahtani:2017jwl,Alqahtani:2017tnq}.  

The structure of our paper is as follows.  In section~\ref{sec:pot} we review how the complex-valued heavy-quark potential used for the description of bottomonium is defined and extracted using lattice QCD and appropriately parametrized for use in phenomenological applictions. Section~\ref{sec:dynamics} collects the relevant details on the dynamical evolution model of anisotropic hydrodynamics and how it is connected to the lattice-vetted potential via a spacetime-dependent anisotropic Debye mass. We present the results of our computation in section~\ref{sec:results} and conclude with a discussion and outlook in section~\ref{sec:con}.


\section{Potential description of heavy-quarkonium}
\label{sec:pot}

Herein, we use a lattice QCD vetted, non-relativistic potential based description of heavy-quarkonium in a thermal medium to compute its real-time evolution in a heavy-ion collision. The potential originates in a systematic treatment of heavy quarkonium in QCD, based on the effective field theories non-relativistic QCD (NRQCD) and potential NRQCD \cite{Brambilla:2004jw,Brambilla:2008cx}. These frameworks exploit the inherent separation of scales between the heavy quark rest mass $m_{\rm c,b}$, the surrounding medium temperature, as well as the characteristic scale of QCD $\Lambda_{\rm QCD}$ to dramatically simplify the description of heavy quarkonium. Instead of having to consider a full quantum field-theoretical boundary value problem for Dirac spinor fields, one may go over in a first step to an initial-value problem for two-component Pauli spinors (NRQCD). In a second step this non-relativistic theory may be matched to a further simplified description in terms of coupled color singlet $ \psi_S({\bf r},t)$ and color octet wavefunctions $ \psi_O({\bf r},t)$ (pNRQCD). In the latter, the interaction among the heavy-quarks, as well as their interaction with the medium is captured in both potential and non-potential contributions. Depending on the concrete scale hierarchy of the problem at hand, the potential contributions may dominate and the relevant real-time evolution of heavy quarkonium reduces to a Schr\"odinger equation. 

For realistic settings, such as those encountered in heavy-ion collisions at RHIC and LHC, the matching coefficients of the effective theory, i.e. the potential cannot be reliably determined using perturbation theory. Nevertheless vital insight had been gained by evaluating pNRQCD using the hard-loop approximation \cite{Laine:2006ns,Brambilla:2008cx}. In particular it was found that  the proper in-medium potential must assume complex values at high temperatures, as was first discussed in \cite{Laine:2006ns} and subsequently extended to a momentum-space anisotropic medium in \cite{Dumitru:2007hy,Burnier:2009yu,Dumitru:2009fy,Nopoush:2017zbu}. This fact, in particular, implies that purely real model potentials, such as the popular color singlet free energies or the internal energies do not constitute valid descriptions of the relevant in-medium quarkonium physics. 

The proper heavy-quark potential is related to a real-time QCD quantity, the rectangular Wilson loop, via the process of matching,~i.e.~one selects a correlation function in the effective theory pNRQCD and in the underlying microscopic theory QCD which carry the same physics content and identifies them at an appropriate scale. In our case the unequal time correlation function of a heavy quarkonium singlet state may be identified with the Wilson loop in the static limit
\begin{eqnarray}
\nonumber \Big\langle \psi_S({\bf r},t)\psi_S({\bf r},0) \Big\rangle_{\rm pNRQCD} \overset{m\to\infty}{\equiv}  W_\square({\bf r},t)=\Big\langle {\rm Tr} \Big( {\rm exp}\Big[-ig\int_\square dx^\mu A_\mu^aT^a\Big] \Big) \Big\rangle_{\rm QCD}.
\label{Eq:ForwProp}
\end{eqnarray}
Since the Wilson loop obeys a simple equation of motion \cite{Laine:2006ns}
\begin{align}
i\partial_tW_\square(r,t)=\Phi(r,t)W_\square(r,t),
\end{align}
with a time- and space-dependent complex function $\Phi(r,t)$,  the potential picture is applicable as long as $\Phi$ asymptotes towards a time independent value at late times. This value in general is complex and the corresponding potential may be formally defined as 
\begin{align}
V_{\rm QCD}(r)=\lim_{t\to\infty} \frac{i\partial_t W(t,r)}{W(t,r)}\label{Eq:VRealTimeDef}.
\end{align}

As such, this real-time definition is not yet amenable to an evaluation in non-perturbative lattice QCD, which is simulated in unphysical Euclidean time. Instead, one has to take a detour via the spectral decomposition of the Wilson loop to relate the Euclidean and Minkowski domain \cite{Rothkopf:2009pk,Rothkopf:2011db}
\begin{align}
 W_\square(\tau,r)=\int d\omega e^{-\omega \tau} \rho_\square(\omega,r)\,
\leftrightarrow\, \int d\omega e^{-i\omega t} \rho_\square(\omega,r)= W_\square(t,r). \label{Eq:SpecDec}
\end{align}
Inserting \eqref{Eq:SpecDec} into \eqref{Eq:VRealTimeDef} tells us that the real and imaginary part of the potential are related to the position and width of the lowest lying peak structure within the Wilson loop spectrum. It has been shown \cite{Burnier:2012az} that if a potential picture is applicable, the Wilson loop spectrum actually contains a well defined lowest lying peak of skewed Lorentzian form from which the values of the potential can be straightforwardly extracted via a $\chi^2$-fit. 

Note however that the extraction of spectral functions from Euclidean lattice data is an ill-posed inverse problem, which has only recently been successfully tackled in the context of the heavy-quark potential. With the help of a novel Bayesian approach \cite{Burnier:2013nla} the reconstruction robustness was significantly improved compared to previous attempts based on the Maximum Entropy Method \cite{Rothkopf:2011db}. In practice, instead of the Wilson loop on the lattice, one considers Wilson line correlators fixed to Coulomb gauge, which are free from a class of divergences hampering the numerical determination of the Wilson loop.  Using this prescription, the values of the potential have been extracted, to date, in quenched QCD based on the naive Wilson action \cite{Burnier:2016mxc}, as well as for full QCD with $N_f=2+1$ light medium quark flavors based on simulations by the HotQCD collaboration. In both cases, the applicability of the potential picture was confirmed, at all temperatures considered, as the Wilson spectral functions showed a well defined peak of Lorentzian shape. 

To utilize the discrete values of the potential obtained from lattice QCD two further steps need to be taken, for which we follow the same strategy as laid out in \cite{Burnier:2015tda,Burnier:2016kqm}. The first task is to parametrize the values of the potential with an analytic formula, which allows the evaluation of Re[V] and Im[V] at intermediate separation distances not explicitly resolved on the lattice. The second task is to correct the parameters of this analytic parametrization for finite volume and finite lattice spacing artifacts, as no fully continuum extrapolated lattice QCD determination of the potential has been achieved so far.

The analytic parametrization we deploy in the following is based on the concept of a generalized Gauss law for the vacuum heavy quark potential. Lattice QCD studies have shown that over the phenomenologically relevant range of distances for quarkonium the $T=0$ potential is very well reproduced by the Cornell ansatz, consisting of a Coulombic and linearly rising term. Allowing the linear term to contribute down to the smallest distances mimics the effects of a running in the coupling. With the Cornell potential applicable in vacuum, we can consider the divergence of  the auxiliary (color) electric field $\vE=q r^{a-1} \rh$ arising from either the the Coulombic $a=-1, q=\alpha_s, [\alpha_s]=1$ or the string-like part $a=1, q=\sigma, [\sigma]={\rm GeV}^2$
\begin{align}
\vec\nabla \left(\frac{\vE}{r^{a+1}}\right)=   4\pi \,q\, \delta(\vr) \label{Eq:GenGauss}.
\end{align}
Three parameters enter this expression, which characterize the non-perturbative vacuum physics of the quarkonium bound state: the strong coupling $\alpha_s$, the string tension $\sigma$, and a constant shift $c$. Note that we absorb the factor $C_F$ into our definition of the strong coupling $\alpha_s=\frac{g^2C_F}{4\pi}$.

In order to introduce the effects of a thermal medium, we adopt a prescription well known in classical electrodynamics,~i.e.~in the case of the Coulombic contribution, one Fourier transforms Gauss' law and subsequently modifies the right hand side by dividing it with an in-medium permittivity $\epsilon$. We use here the permittivity of the QCD medium computed in hard-thermal loop perturbation theory  
\begin{align}
\eps^{-1}(\vp,m_D)=\frac{p^2}{p^2+m_D^2}-i\pi T \frac{p\, m_D^2}{(p^2+m_D^2)^2}.\label{Eq:HTLperm}
\end{align}
The idea is that the non-perturbative physics of the bound state is encoded in the Cornell form of the $T=0$ potential, whose modification is driven by a weakly-coupled gas of quarks and gluons. Combining \eqref{Eq:HTLperm} and \eqref{Eq:GenGauss} thus leads to integro-differential equations for the in-medium modified Coulombic and string-part of the $T=0$ potential as discussed in detail in \cite{Burnier:2015nsa}. Since the permittivity is complex, the in-medium potential also contains an imaginary part. In contrast to purely perturbative computations, which capture only the Coulombic contribution to the potential, the in-medium potential here receives additional contributions to its real and imaginary part from the string-like part of the $T=0$ Cornell potential. The explicit expressions for the Coulombic part are
\begin{align}
V_c(r)= -\alpha_s\left[\mD+\frac{e^{-\mD r}}{r}
+iT\phi(\mD r)\right] \label{Eq:VHTL},
\end{align}
with 
\begin{align}
\phi(x)=2 \int_0^\infty dz \frac{z}{(z^2+1)^2}\left(1-\frac{\sin(xz)}{xz}\right),\label{phi}
\end{align}
which coincide with the results of Ref.~\cite{Laine:2006ns}.
The additional and novel string-like contribution on the other hand reads
\begin{align}
{\rm Re}V_s(r)&=-\frac{\Gamma[\frac{1}{4}]}{2^{\frac{3}{4}}\sqrt{\pi}}\frac{\sigma}{\mu} D_{-\frac{1}{2}}\big(\sqrt{2}\mu r\big)+ \frac{\Gamma[\frac{1}{4}]}{2\Gamma[\frac{3}{4}]} \frac{\sigma}{\mu} \, ,
\end{align}
for the real part, where the strength of the in-medium modification is characterized by the parameter $\mu^4=m_D^2 \sigma/\alpha_s$. For its imaginary part we have 
\begin{eqnarray}
\Im V_s(r)&=&-i\frac{\sigma m_D^2 T}{\mu^4} \psi(\mu r)=-i\alpha_s T \psi(\mu r)\label{Eq:ImVSGenGauss},
\end{eqnarray}
where $\psi$ corresponds to the following Wronskian
\begin{eqnarray} 
 \notag\psi(x)&=&D_{-1/2}(\sqrt{2}x)\int_0^x dy\, {\rm Re}D_{-1/2}(i\sqrt{2}y)y^2 \phi(ym_D/\mu)\\\notag&&\hspace{-4mm}+{\rm Re}D_{-1/2}(i\sqrt{2}x)\int_x^\infty dy\, D_{-1/2}(\sqrt{2}y)y^2 \phi(ym_D/\mu)\\ \nonumber&&-D_{-1/2}(0)\int_0^\infty dy\, D_{-1/2}(\sqrt{2}y)y^2 \phi(ym_D/\mu).
\end{eqnarray}
An important aspect of these expressions is that, once the vacuum parameters of the Cornell potential are fixed, only a single temperature-dependent parameter remains, the Debye mass $m_D$.

\begin{figure*}
\includegraphics[scale=0.4]{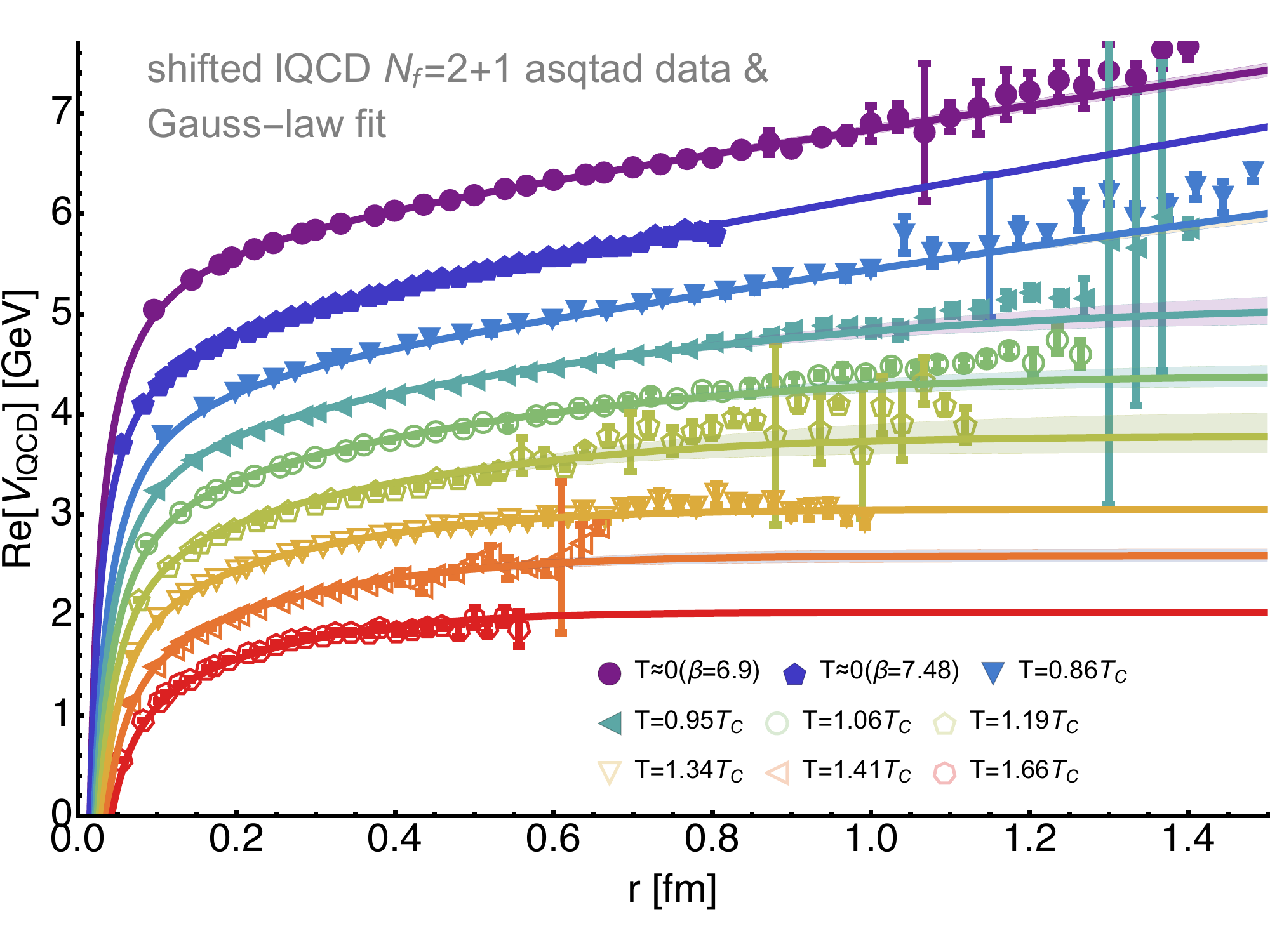}
\includegraphics[scale=0.4]{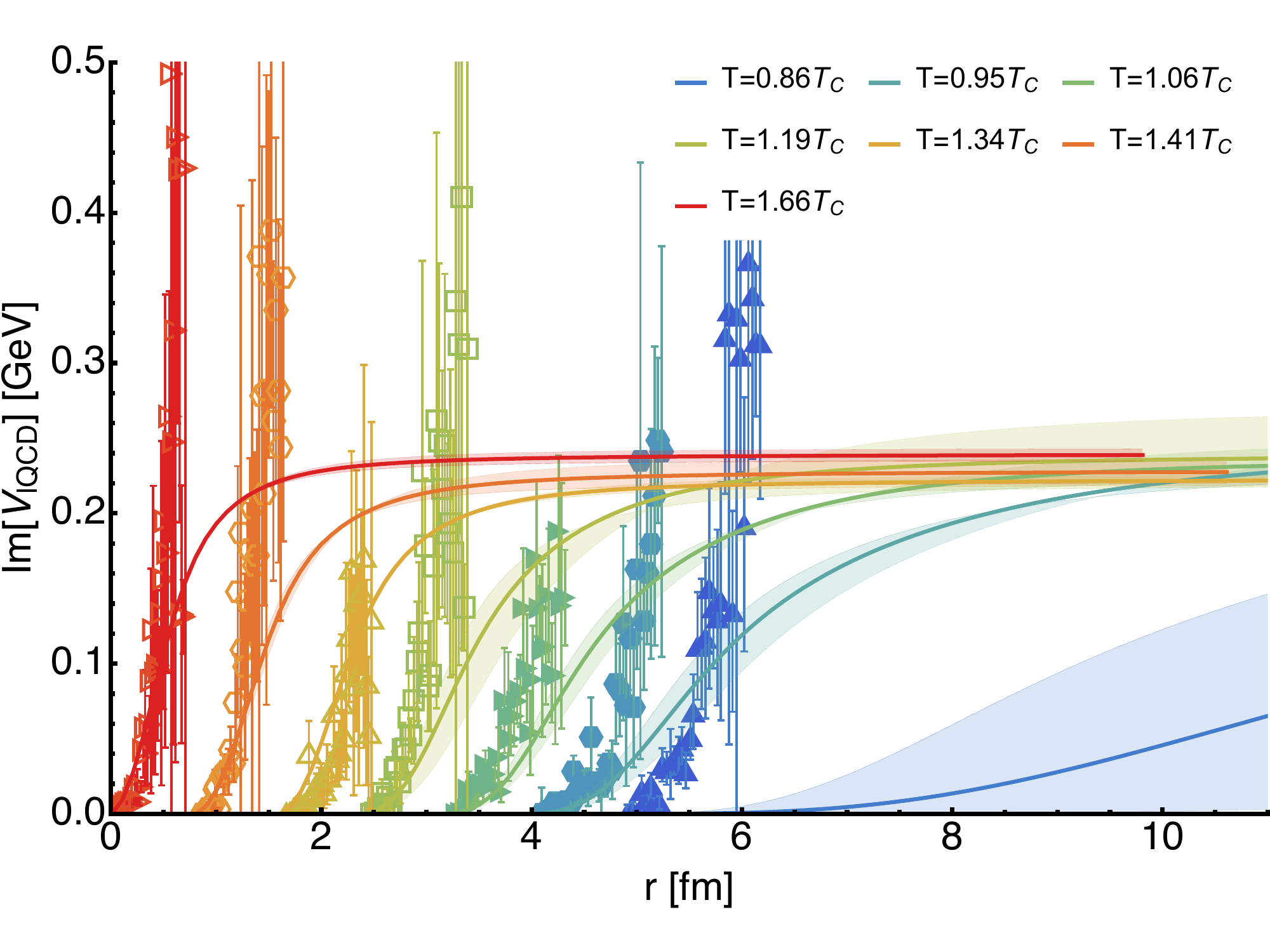}
\caption{The real (left) and imaginary (right) part of the in-medium heavy quark potential in full QCD with $N_f=2+1$ light quark flavors based on ensembles by the HotQCD collaboration (colored points, shifted for better readability). The vacuum parameters here are tuned using the $T\approx0$ ensembles at $\beta=6.9$ and $\beta=7.48$. By adjusting the Debye mass parameter $m_D$, the lattice QCD values of $\Re[V]$ are reproduced very well via the Gauss-law parametrization (solid lines) over all separation distances and temperatures. The theoretical error bars (shown as shaded regions surrounding the central line) arise from the fit uncertainty of $m_D$. For $\Im[V]$, the agreement at high temperatures and small distances is very good, while at $T\approx T_C$ deviations from the extracted lattice values are visible. (The crossover temperature on these lattices due to the relatively large pion mass of $m_\pi\approx300$MeV lies at $T_C=172.5$ MeV) }\label{Fig:PotentialLat}
\end{figure*}

While the analytic parametrization was derived in a straightforward fashion, it relies on several assumptions and thus needs to be validated on real lattice QCD data. Using both quenched \cite{Burnier:2016mxc} and full QCD simulations with $N_f=2+1$ light flavors \cite{Burnier:2015tda} it was shown that the lattice values of the potential were indeed excellently reproduced by the generalized Gauss law parametrization (see Fig.\ref{Fig:PotentialLat}). After fixing $\alpha_s$, $\sigma$, and $c$ using low temperature ensembles, the real-part of the potential was fitted by tuning $m_D$. Once $m_D$ is fixed, the parametrization makes a prediction for $\Im[V]$, which for quenched QCD simulations showed quantitative agreement at high temperatures and, as expected, became less accurate around the phase transition. In full QCD, no robust determination of the imaginary part has been achieved so far, however, the tentative values extracted, again, showed very good agreement with the Gauss-law parametrization at high and intermediate temperatures \cite{Burnier:2015tda}. The values for the Debye mass related to the full QCD in-medium potential also showed clear deviations from the perturbative predictions in the phenomenologically relevant regime between $T_C<T<3T_C$.

Even though it has been established that the Gauss-law parametrization is capable of reproducing the lattice QCD in-medium potential, its values found on the lattice may not be applied directly to phenomenological computations due to the presence of lattice artifacts. If a continuum extrapolation of the $T=0$ potential in the thermodynamic limit were available, we could directly determine the parameters $\alpha_s$, $\sigma$ and $c$ from first principles. Here instead we select these parameters in a phenomenological fashion, i.e. we tune them such that the vacuum bottomonium spectrum below the B-meson threshold is reproduced. The correct quark mass to be used in such a computation is the renormalon subtracted mass, which for bottomonium may be perturbatively computed and takes the value $m^{\rm RS'}_b=4.882\pm0.041$GeV. Since the vacuum potential in full QCD was robustly determined only up to distances \mbox{$r\approx1$ fm} we enforce the flat asymptotics due to string breaking by hand at $r_{\rm SB}=1.25$ fm. Within this setting the best set of parameters is
\begin{align}
c=-0.1767\pm 0.0210~{\rm GeV}, \quad \alpha_s=0.5043\pm 0.0298,\quad \sqrt{\sigma}=0.415\pm0.015~{\rm GeV}.\label{T0const}
\end{align}

\begin{figure*}
\includegraphics[scale=0.4]{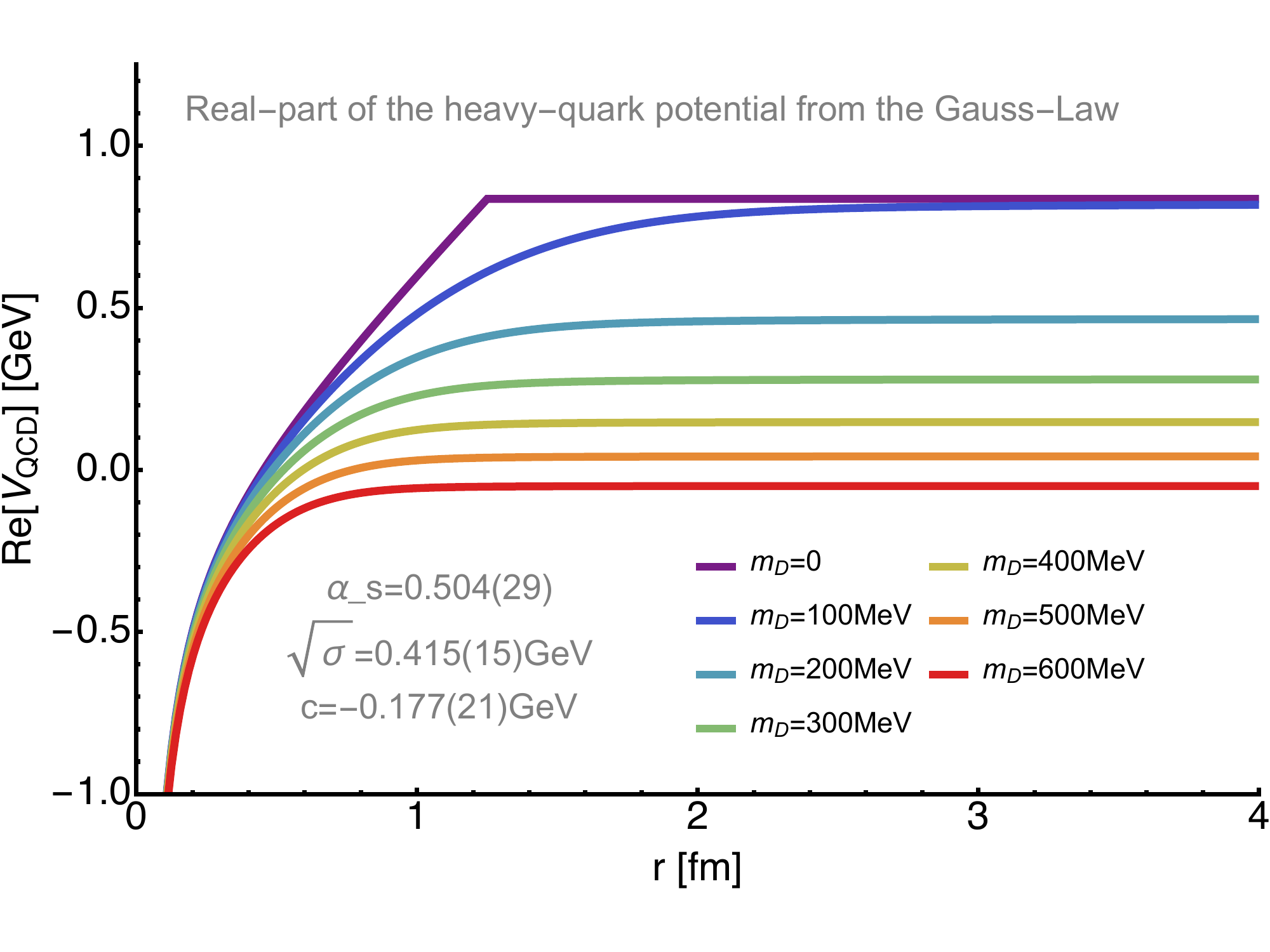}
\includegraphics[scale=0.4]{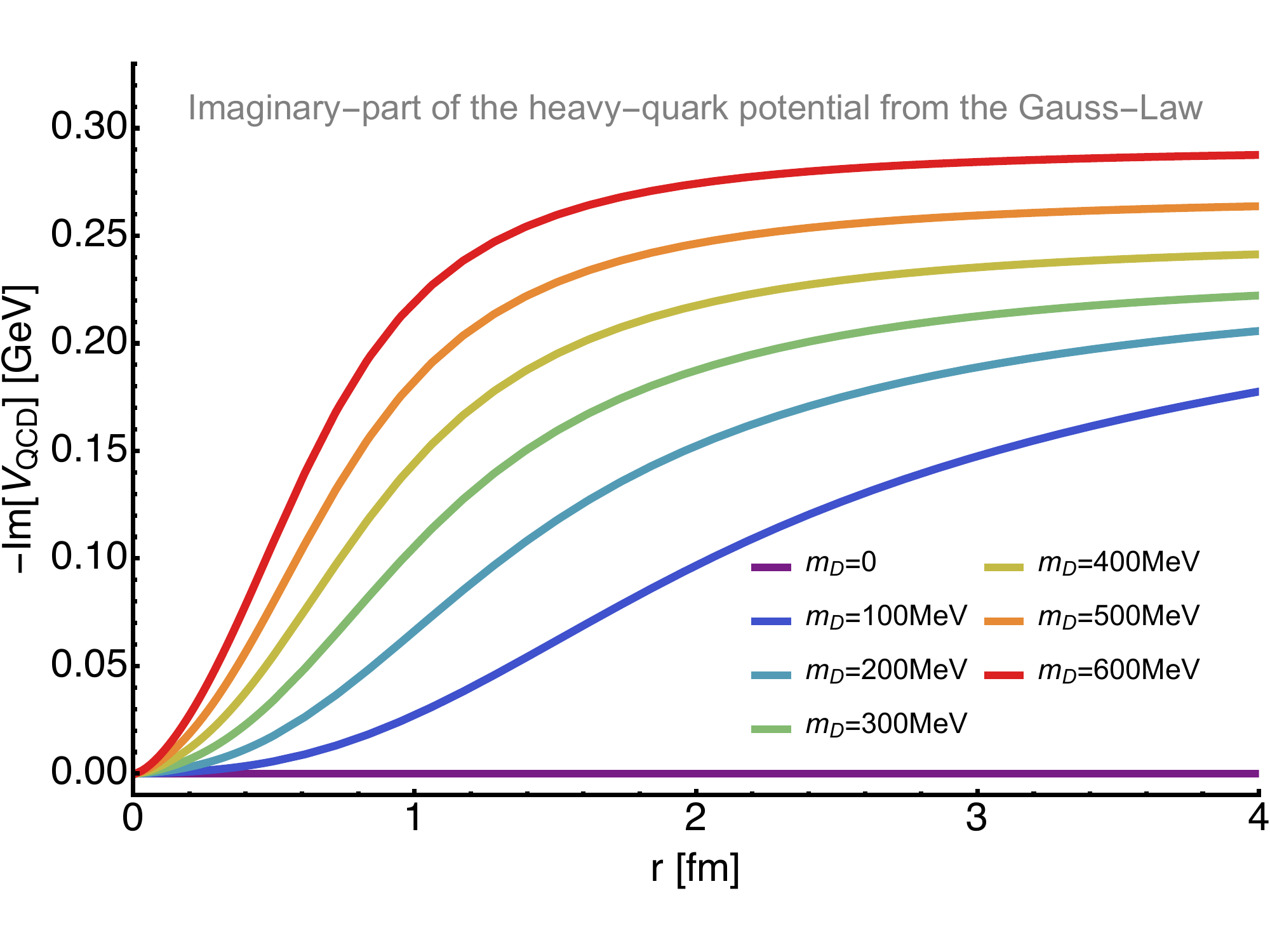}
\caption{The real (left) and imaginary (right) part of the in-medium heavy quark potential used in this study. Their values are given for different values of the Debye mass of the QCD medium. The vacuum parameters $\alpha_S$, $\sigma$, and $c$ at $m_D=0$ were tuned such that the PDG bottomonium spectrum is reproduced. To this end string breaking is enforced at $r_{\rm sb}=1.25$ fm. The in-medium modification of the Cornell-type vacuum potential in our approach is governed by a single temperature-dependent parameter, the Debye mass $m_D$. Thermal effects lead to a characteristic (Debye) screening of the real-part and induce a finite imaginary part, which saturates at large distances.}\label{Fig:Potential}
\end{figure*}

In this study we will compute the Debye mass self-consistently from the dynamical evolution of the bulk and use its value to implement the in-medium modification of the Cornell potential with the above parameters. In Fig.~\ref{Fig:Potential} we show the real (left) and imaginary part (right) of the actual potential used for different values of the Debye mass. In order understand better which values of $m_D$ play a role in the evolution of heavy quarkonium we note that lattice QCD studies showed that in a thermal QCD medium close to the crossover transition the ratio of $m_D/T\approx1$ and grows to $m_D/T\approx2$ as temperature is increased to $T=2T_C$.



\section{Non-equilibrium corrections, static Schr\"odinger equation solution, and real-time QGP evolution}
\label{sec:dynamics}

With the real and imaginary parts of the potential specified, we can now turn to the method used to fold the dynamical evolution of the full three-dimensional QGP evolution together with information about the real and imaginary parts of the resulting binding energies.  In order to do so, however, we will must first make an extension of the result presented in the previous section to the case of a QGP with momentum-space anisotropies. This is necessary for a realistic model of QGP evolution and quarkonia suppression.  The simplest form for the soft-particle distribution function that can be used to take into account QGP momentum-space anisotropies is a generalization of an isotropic phase-space distribution which is squeezed or stretched along one direction in momentum space, defined by $\hat{\bf n}$, with a parameter $-1 < \xi < \infty$.  In a heavy-ion collision the direction $\hat{\bf n}$ can be identified with the beam line direction, $\hat{\bf n} = \hat{\bf z}$.  The resulting one-particle distribution function is given by the following spheroidal ``Romatschke-Strickland'' form \cite{Romatschke:2003ms,Romatschke:2004jh}
\begin{equation}
f({\bf p},\xi,\Lambda)\equiv f_{\rm iso}(\sqrt{{\bf p}^2+\xi({\bf p\cdot \hat{n}})^2}/\Lambda) \, ,
\label{distansatz}
\end{equation}
where $\Lambda$ is the transverse temperature scale and $f_{\rm iso}$ is the isotropic thermal distribution function associated with the soft degrees of freedom in the QGP.

\subsection{Anisotropic Debye mass}
\label{ssec:amd}

The Debye mass, which we use to evaluate the heavy-quark potential in our approach, is self consistently computed from the dynamical evolution of the soft bulk matter. The conventional isotropic $m_D$ is defined via an integral over the isotropic distribution function
\be \label{eq:Debye}
m_D^2 = -\frac{g^2}{2\pi^2} \int_0^\infty d p \,
  p^2 \, \frac{d f_{\rm iso}}{d p} ~,
\ee
where $p^2 \equiv {\bf p}^2 = p_\perp^2 +p_z^2$. Since lattice studies are restricted to a thermal and isotropic state, we have to use additional input to incorporate the effects of a QGP momentum-space anisotropy on the potential itself.  For this purpose, we use the results contained in Sec.~2.2 of Ref.~\cite{Strickland:2011aa} where it was shown that, empirically, one can incorporate the momentum-space anisotropy parameter $\xi$ into the potential by making the Debye mass depend on the angle with respect to the beam line direction.  The result obtained was
\be
\left(\frac{\mu}{m_D}\right)^{-4} =  
1 + \xi\left(a - \frac{2^b(a-1)+(1+\xi)^{1/8}}{(3+\xi)^b}\right) 
\left(1 + \frac{c(\theta) (1+\xi)^d}{(1+ e\xi^2)} \right) \, ,
\label{eq:muparam}
\ee
with $a=16/\pi^2$, $b = 1/2$, $d=3/2$, $e=1/3$, and
\be
c(\theta) = \frac{3 \pi ^2 \cos (2 \theta)+\left(9+4 \sqrt{3}-4 \sqrt{6}\right) \pi ^2+64 \left(\sqrt{6}-3\right)}{4
   \left(\sqrt{3} \left(\sqrt{2}-1\right) \pi ^2-16 \left(\sqrt{6}-3\right)\right)} \, .
\ee
With these values in hand, we simply tabulate the real and imaginary parts of the potential functions with respect to $m_D$ and $r$, and then replace $m_D \rightarrow \mu$ in the isotropic potential to obtain the corresponding anisotropically modified potential. The parameter $\xi$ sets the degree of momentum-space anisotropy, with $-1 < \xi < 0$ corresponding to a prolate distribution, with more momentum along the beam line direction than transverse to it and $\xi >0$ corresponding to an oblate distribution.  The anisotropy parameter is dynamically tied to the bulk QGP evolution using anisotropic hydrodynamics (aHydro) which we will discuss shortly \cite{Martinez:2010sc,Florkowski:2010cf,Martinez:2010sd,Ryblewski:2010bs,Ryblewski:2011aq,Martinez:2012tu,Ryblewski:2012rr,Strickland:2014pga,Nopoush:2014pfa,Alqahtani:2015qja,Alqahtani:2016rth,Alqahtani:2017jwl,Alqahtani:2017tnq}.

\subsection{Solving the Schr\"odinger equation with a complex potential}
\label{ssec:schrod}

Since the potential no longer has full rotational symmetry one has to go beyond a single (radial) dimension when solving the Schr\"odinger equation.  With the assumed form for the one-particle distribution function, only one symmetry direction is broken (spheroidal symmetry), and one can, in principle, simply use a two-dimensional solver. Here we have chosen to make use of a full three-dimensional solution since codes for this are already available even for complex-valued potentials \cite{Strickland:2009ft,Margotta:2011ta}.  For this purpose, the real-time Schr\"odinger equation is solved in imaginary time. The grid spacing for $\Upsilon(nS)$ states is chosen as $a=0.15$ fm on $N=256^3$ regularly-spaced grid points. Due to the larger size of p-wave bottomonia, our grid spacing for the $\chi_{b}(mP)$ states is set to $a=0.175$ fm. Starting from a randomized three-dimensional wavefunction, we evolve in imaginary time until the ground state wavefunction converges to within a given tolerance.  Using ``snapshots'' of the wavefunction stored during the imaginary-time evolution, we can project out the low-lying excited states \cite{Strickland:2009ft}.  Additionally, by fixing the symmetry (symmetric vs anti-symmetric) of the initial random wavefunction we can select the s-wave or p-wave states independently \cite{Strickland:2009ft}.  In this way we can obtain the wavefunctions of the 1s, 2s, 3s, 1p, and 2p Upsilon states and in turn we can compute the real and imaginary parts of their respective energies. For more details concerning the numerical method, we refer the reader to Refs.~\cite{Strickland:2009ft,Margotta:2011ta}.

\subsection{Anisotropic hydrodynamics equations and initial conditions}
\label{ssec:3p1}

To proceed, we assume that the underlying bulk one-particle distribution function is well approximated by Eq.~(\ref{distansatz}) at all points in spacetime. In addition, we assume that the bulk evolution is well described using hydrodynamical degrees of freedom, such as energy density, pressures, and viscous corrections.  This assumption has been tested by comparing the predictions of anisotropic hydrodynamics to exact solutions of the Boltzmann equation~\cite{Florkowski:2013lza,Florkowski:2013lya,Bazow:2013ifa,Florkowski:2014sfa,Florkowski:2014sda,Denicol:2014xca,Denicol:2014tha,Nopoush:2014qba,Molnar:2016gwq,Martinez:2017ibh} (see also \cite{Damodaran:2017ior} for more comparisons to kinetic theory) where it has been shown to reproduce the exact kinetic evolution even far from equilibrium.  The equations used herein are obtained using the zeroth and first moments of the Boltzmann equation in the relaxation-time approximation.  For details about the dynamical equations used and their physical content, we refer the reader to Sec.~IV of Ref.~\cite{Ryblewski:2015hea}.

In order to solve the aHydro dynamical equations, one has to make a reasonable assumption about the initial conditions at the initial longitudinal proper-time for the hydrodynamic evolution, $\tau = \tau_0$.  For our initial conditions, we take the system to be isotropic in momentum space ($\xi=0$) with zero transverse flow and Bjorken flow in the longitudinal direction.  In the transverse plane, the initial energy density is computed from a linear combination of smooth Glauber wounded-nucleon and binary-collision profiles with a binary mixing factor of $\alpha = 0.15$.  In the longitudinal direction, we used a ``tilted'' profile with a central plateau and Gaussian tails resulting in a profile function of the form $\rho(\varsigma) \equiv \exp \left[ - (\varsigma - \Delta \varsigma)^2/(2 \sigma_\varsigma^2) \, \Theta (|\varsigma| - \Delta \varsigma) \right]$, with \mbox{$\varsigma = \arctanh(z/t)$} being spatial rapidity.  The parameters entering the longitudinal profile function were fitted to the pseudorapidity distribution of charged hadrons with the results being $\Delta\varsigma = 2.3$ and $\sigma_{\varsigma} = 1.6$.  The first quantity sets the width of the central plateau and the second sets the width of the Gaussian ``wings''.  The resulting initial energy density at a given transverse position ${\bf x}_\perp$ and spatial rapidity $\varsigma$ was computed using ${\cal E} \propto (1-\alpha) \rho(\varsigma) \left[ W_A({\bf x}_\perp) g(\varsigma) + W_B({\bf x}_\perp) g(-\varsigma)\right] + \alpha \rho(\varsigma) C({\bf x}_\perp)$, where $W_{A,B}({\bf x}_\perp)$ is the wounded nucleon density for nuclei $A$ and $B$, $C({\bf x}_\perp)$ is the binary collision density, and $g(\varsigma)$ is the ``tilt function''.  The tilt function $g(\varsigma) = 0$ if $\varsigma < -y_N$, $g(\varsigma) = (\varsigma+y_N)/(2y_N)$ if $-y_N \leq \varsigma \leq y_N$, and $g(\varsigma)=1$ if $\varsigma > y_N$ where $y_N = \log(2\sqrt{s_{NN}}/(m_p + m_n))$ is the nucleon momentum rapidity \cite{Bozek:2010bi}.

\subsection{Computing the suppression}
\label{ssec:sup}

The solution of the 3+1d aHydro dynamical equations gives us hard momentum scale $\Lambda$, the momentum-anisotropy parameter $\xi$, and, consequently, the anisotropic Debye mass $\mu$ as a function of proper time, transverse coordinate 
${\bf x}_\perp$, and spatial rapidity $\varsigma$.  Solving the Schr\"odinger equation in addition provides the real and 
imaginary parts of the binding energy of a given state as a function of $\Lambda$ and $\xi$.  Folding these together 
gives us the real and imaginary parts of the binding energy 
as a function of proper time, transverse coordinate ${\bf x}_\perp$, and 
spatial rapidity $\varsigma$: $\Re[E_{\rm bind}(\tau,{\bf x}_\perp,\varsigma)]$ and 
$\Im[E_{\rm bind}(\tau,{\bf x}_\perp,\varsigma)]$, respectively.  Full details of the method used can be found in Ref.~\cite{Strickland:2011aa}.  Here we summarize the important points.

We use the imaginary part of the binding energy to provide information about the decay rate of a given state and the real part of the binding energy to tell us when the state becomes completely unbound.  The imaginary part of the binding energy can be related to the decay rate, $\Gamma$, using $\partial_\tau n_i = 
-\Gamma_i n_i$, where $i$ indexes the state in question, giving $\Gamma_i = -2 \Im[E_i]$ \cite{Strickland:2011aa}.
Finally, using the fact $\Im[E_{\rm bind}] = - \Im[E]$ we obtain 
\be
\Gamma(\tau,{\bf x}_\perp,\varsigma) = 
\left\{
\begin{array}{ll}
2 \Im[E_{\rm bind}(\tau,{\bf x}_\perp,\varsigma)]  & \;\;\;\;\; \Re[E_{\rm bind}(\tau,{\bf x}_\perp,\varsigma)] >0 \\
10\;{\rm GeV}  & \;\;\;\;\; \Re[E_{\rm bind}(\tau,{\bf x}_\perp,\varsigma)] \le 0 \\
\end{array}
\right.
\ee
The value of 10 GeV in the second case was chosen in order to quickly suppress states which are
fully unbound and, in practice, the results do not depend significantly on this value as long as it is large enough to quickly disassociate the state under consideration.

We then integrate the instantaneous decay rate, $\Gamma$, obtained in this manner
over proper-time to extract the dimensionless logarithmic suppression factor 
\be
\zeta(p_T,{\bf x}_\perp,\varsigma) \equiv \Theta(\tau_f-\tau_{\rm form}(p_T)) \int_{{\rm max}(\tau_{\rm form}(p_T),\tau_0)}^{\tau_f} 
d\tau\,\Gamma(\tau,{\bf x}_\perp,\varsigma) \, ,
\label{eq:zeta}
\ee
where $\tau_{\rm form}(p_T)$ is the lab-frame formation time of the state in question.
The formation time of a state in its local rest frame 
can be estimated by the inverse of its vacuum binding energy~\cite{Karsch:1987uk}.
For the formation times for the $\Upsilon(1S)$, $\Upsilon(2S)$, $\Upsilon(3S)$, $\chi_{b}(1P)$, $\chi_{b}(2P)$, $\chi_{b}(3P)$ and states we take $\tau_{\rm form}^0$ = 0.2 fm/c, 0.4 fm/c, 0.6 fm/c, 0.4 fm/c, 0.6 fm/c, and 0.6 fm/c, respectively.

Our choice for the initial proper time $\tau_0$ for plasma evolution is $\tau_0 =$ 0.3 fm/c at both RHIC and LHC energies.
The final time, $\tau_f$, is defined to be the proper time when the local effective temperature drops below $T_d = 192$ MeV. At this energy density, plasma screening effects are assumed to decrease
rapidly due to the transition to the hadronic phase and the widths of the states will become
approximately equal to their vacuum widths.  

From $\zeta$ obtained via Eq.~(\ref{eq:zeta}) one can directly compute the suppression factor $R_{AA}$ using
\be
R_{AA}(p_T,{\bf x}_\perp,\varsigma) = e^{-\zeta(p_T,{\bf x}_\perp,\varsigma)} \, .
\ee
Next, we must average over transverse momenta, implementing the appropriate cuts.
For this purpose, we assume that 
all states have an approximately $1/E_T^4$ spectrum.  Integrating over transverse momentum given $p_T$-cuts $p_{T,\rm min}$ and
$p_{T,\rm max}$ we obtain the $p_T$-cut suppression factor
\be
R_{AA}({\bf x}_\perp,\varsigma) \equiv \frac{\int_{p_{T,\rm min}}^{p_{T,\rm max}} 
dp_T^2 \, R_{AA}(p_T,{\bf x}_\perp,\varsigma) /(p_T^2 + M^2)^2}{\int_{p_{T,\rm min}}^{p_{T,\rm max}}dp_T^2/(p_T^2 + M^2)^2} \, .
\ee
For implementing cuts in centrality we compute $R_{AA}$ for finite impact parameter $b$ and map centrality
to impact parameter in the standard manner. Finally, to compare with experimental observations we average $R_{AA}({\bf x}_\perp,\varsigma)$
over ${\bf x}_\perp$.  For this operation, we use a production probability distribution which proportional to the overlap density~\cite{Strickland:2011aa}
\be
\langle R_{AA}(\varsigma) \rangle \equiv 
\frac{\int_{{\bf x}_\perp} \! d{\bf x}_\perp \, n_{AA}({\bf x}_\perp)\,%
R_{AA}({\bf x}_\perp,\varsigma)}{\int_{{\bf x}_\perp} \! d{\bf x}_\perp \,%
n_{AA}({\bf x}_\perp)} \, .
\label{eq:geoaverage}
\ee


\section{Results}
\label{sec:results}

In this section, we present our numerical results and compare them to data from RHIC and LHC. The background dynamics discussed in the previous section have been determined for three different values of the shear viscosity to entropy ratio varied around the proposed quantum bound $\eta/s = 1/4\pi$, with the initial temperature tuned so that the soft particle multiplicity is held fixed. The corresponding initial temperature values for the appropriate beam energies are listed in Table \ref{table1} and lead to hydrodynamic evolution, which reproduces light particle spectra and azimuthal flow well. 

\begin{table}[t!]
\begin{tabular}{|c|c|c|c|}
\hline
\backslashbox{$\;4\pi\eta/s$\Big.}{\Big.$\sqrt{s_{NN}}\;$} & \ 0.2 TeV\ \ & \  2.76 TeV \ \ & \  5.02 TeV\ \  \\
\hline
\hline
1 & 0.442 & 0.552 & 0.641\\
\hline
2 & 0.440 & 0.546 & 0.632\\
\hline
3 & 0.439 & 0.544 & 0.629\\
\hline
\end{tabular}
\caption{Values of initial temperature $T_0$ in GeV used for RHIC $0.2$ TeV, LHC run1 $2.76$ TeV, and LHC run2 $5.02$ TeV for the different values of the shear viscosity over entropy ratio considered herein.  In the aHydro simulations used, the QGP was assumed to be initially isotropic in momentum space.}
\label{table1}
\end{table}

\begin{figure*}[t!]
\includegraphics[width=0.55\linewidth]{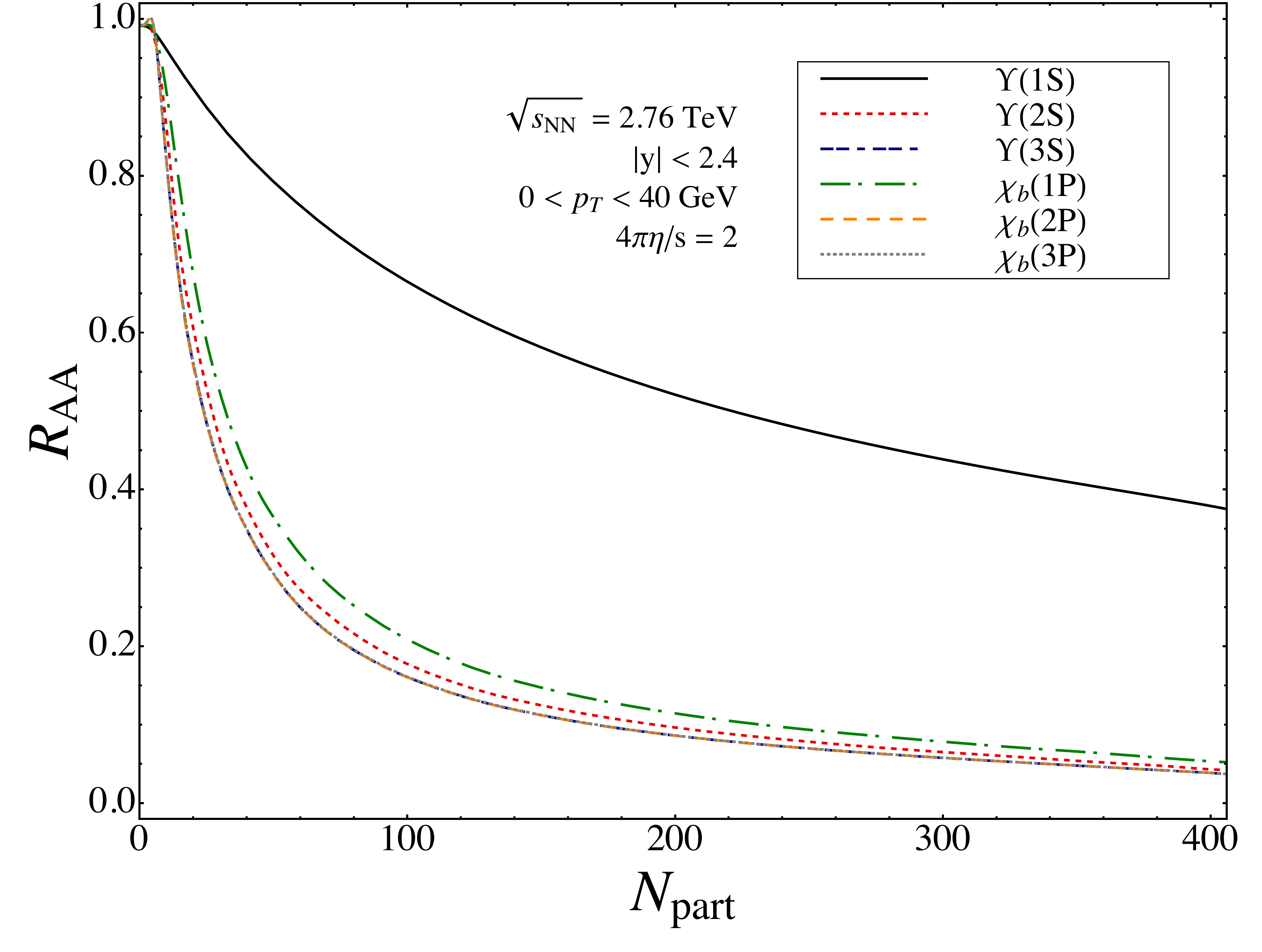}
\caption{
(Color online) Primordial $R_{AA}$ for each bottomonium state as a function of the number of participants for the parameter sets of LHC run1.
}
\label{fig:rawRK}
\end{figure*}

The survival probabilities of the bottomonium states are folded with the aHydro background evolution to obtain the primordial $R_{AA}$ for each state. The resulting values for LHC run1 are shown in Fig. \ref{fig:rawRK}. Due to the large suppression from the lattice-vetted heavy-quark potential, the curves for the $\chi_{b}(2P)$, $\Upsilon(3S)$, and $\chi_{b}(3P)$ states fall on top of each other.  For all of these states, they are completly suppressed in the interior of the fireball, however, there are always states at the edge of the plasma where the temperature/density is low where the states are largely unsuppressed.  This results in a kind of universal ``halo" survival probability which is related to the geometry of the fireball and its temperature profile.

Since it is not the primordial bottomonium states themselves that are measured in the detector but instead the decay di-leptons, feed down must be factored to calculate the inclusive $R_{AA}$ for each state observed in experiment as described in the previous section. The relevant feed down fractions are obtained from averaging over $p_{T}$ from experimental feed down yields \cite{Woeri:2015hq} and are listed in Table \ref{feeddown}.

\begin{table}[t!]
\begin{tabular}{|c|c|c|c|}
\hline
\multicolumn{4}{|c|}{Feed down fractions}\\
\hline
\hline
$\Upsilon(1S) \rightarrow \Upsilon(1S)$ & 0.668 & - & -\\
\hline
$\Upsilon(2S) \rightarrow \Upsilon(1S)$ & 0.086 & $\Upsilon(2S) \rightarrow \Upsilon(2S)$ & 0.604\\
\hline
$\Upsilon(3S) \rightarrow \Upsilon(1S)$ & 0.010 & $\Upsilon(3S) \rightarrow \Upsilon(2S)$ & 0.043\\
\hline
$\chi_{b}(1P) \rightarrow \Upsilon(1S)$ & 0.170 & - & -\\
\hline
$\chi_{b}(2P) \rightarrow \Upsilon(1S)$ & 0.051 & $\chi_{b}(2P) \rightarrow \Upsilon(2S)$ & 0.309\\
\hline
$\chi_{b}(3P) \rightarrow \Upsilon(1S)$ & 0.015 & $\chi_{b}(3P) \rightarrow \Upsilon(2S)$ & 0.044\\
\hline
\end{tabular}
\caption{Feed down fractions to the $\Upsilon(1S)$ state used in the determination of the final measured yields.}
\label{feeddown}
\end{table}

\begin{figure*}[h]
\centerline{
\includegraphics[width=0.65\linewidth]{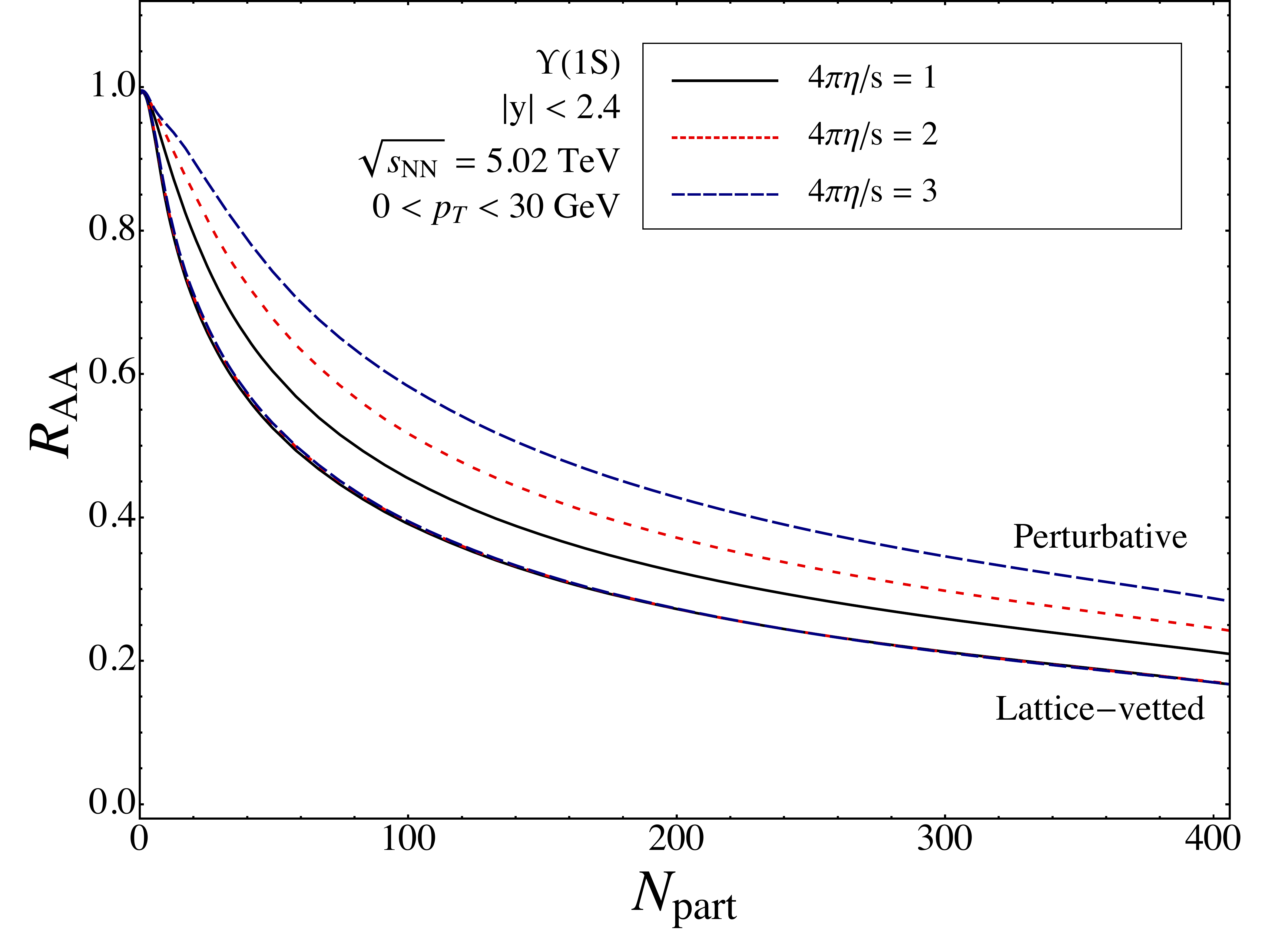}
}
\caption{
(Color online) bottomonium suppression for the LHC run2 parameter set based on the perturbative model of \cite{Strickland:2009ft} (top three curves) and the lattice-vetted heavy-quark potential (bottom three curves).
Note that due to the large imaginary part in the lattice-vetted potential the suppression is consistently larger in than in the perturbative model potential. Interestingly we find that virtually no dependence on the chosen values for the shear viscosity to entropy ratio is observed with the lattice-vetted potential. This behavior is found both at RHIC and LHC energies.
}
\label{fig:inclusivecompare}
\end{figure*}

The inclusive $\Upsilon(1S)$ $R_{AA}$ is presented in Fig. \ref{fig:inclusivecompare} both computed using the lattice-vetted heavy-quark potential (solid lines) as well as with the perturbative Bazow-Strickland model potential of Ref.\cite{Strickland:2011aa} (dashed lines). Two observations can be immediately made: First, the $R_{AA}$ obtained with the lattice-vetted potential lies consistently below the values obtained from the perturbative potential. The reason is that the former features a stronger imaginary part and thus the bottomonium states are more easily dissociated. Secondly we find that that $R_{AA}$ computed with the lattice-vetted heavy-quark potential is virtually independent of the $\eta/s$ parameter of the aHydro background evolution. This behavior is consistently observed both at RHIC and LHC energies and has important consequences for the role bottomonium can play as a probe of the QGP. The less the suppression depends on parameters other than the temperature, the more bottomonium can be used as a as dynamical thermometer of nuclear matter under extreme conditions.

In order to meaningfully compare to experimental results, we need to quantify the uncertainty in the used potential. When vetting the potential with lattice QCD simulations it was observed that the lattice potential values could be fitted with a Debye mass with around 10-20\% error. We, therefore, have repeated our calculations including a modest $\pm 15\%$ variation of $m_{D}$ which leads to an error estimate on the $R_{AA}$ theory curve quantifying the potential uncertainty. Since our results are essentially independent of the shear viscosity parameter, we set $4\pi\eta/s=2$ consistent with recent particle spectra fits \cite{Alqahtani:2017jwl}. 

\begin{figure*}[h]
\includegraphics[width=0.55\linewidth]{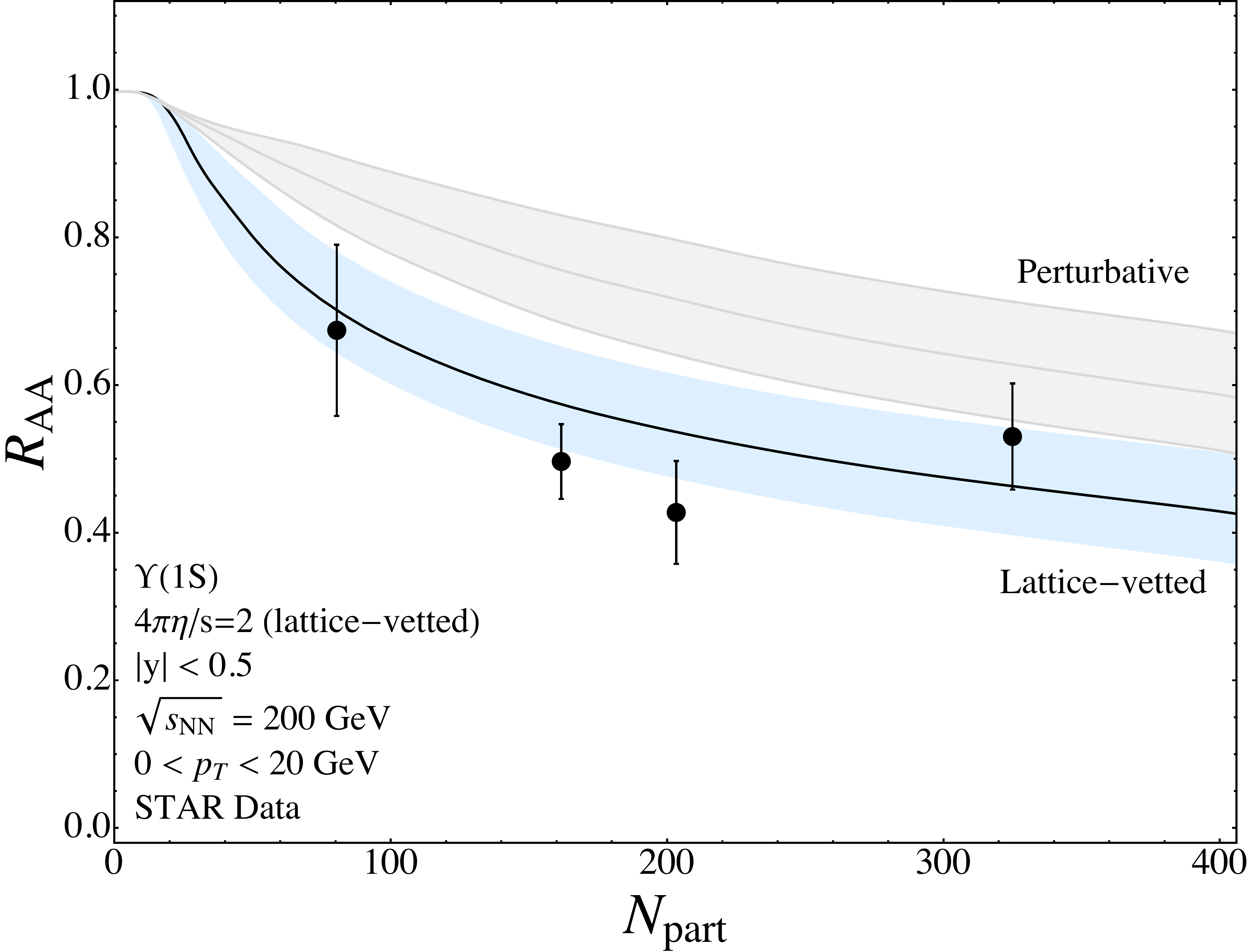}
\caption{
(Color online) $R_{AA}$ as a function of the number of participants compared to STAR data taken at RHIC as a function of $N_{\text{part}}$ for 0.2 TeV Au-Au collisions. We show both the previous estimates based on a model potential (light gray lines with grey error band) and our new results obtained using the lattice-vetted heavy-quark potential (solid line with blue error band). The error-band around our new central value corresponds to a $\pm 15\%$ variation in $m_{D}$ used to estimate the uncertainty in the determination of the potential. Due to the stronger imaginary part present in the lattice-vetted potential, the new estimates move to lower values and are in very good agreement with experimental observations.  
}
\label{fig:STAR}
\end{figure*}

We start the explicit comparisons of our computed yields to experiment with data obtained at RHIC by the STAR collaboration in Fig.~\ref{fig:STAR}. We show both the previous estimates based on a perturbative model potential (light gray lines with grey error band) and based on the lattice-vetted heavy quark potential (solid line with blue error band). As is expected from the behavior found in Fig.~\ref{fig:inclusivecompare} our new results lie systematically below those coming from the perturbative model potential. Interestingly, the shift to lower values brings the values for $R_{AA}$ into very good agreement with the measured RHIC data.

\begin{figure*}[h]
\centerline{
\includegraphics[width=0.47\linewidth]{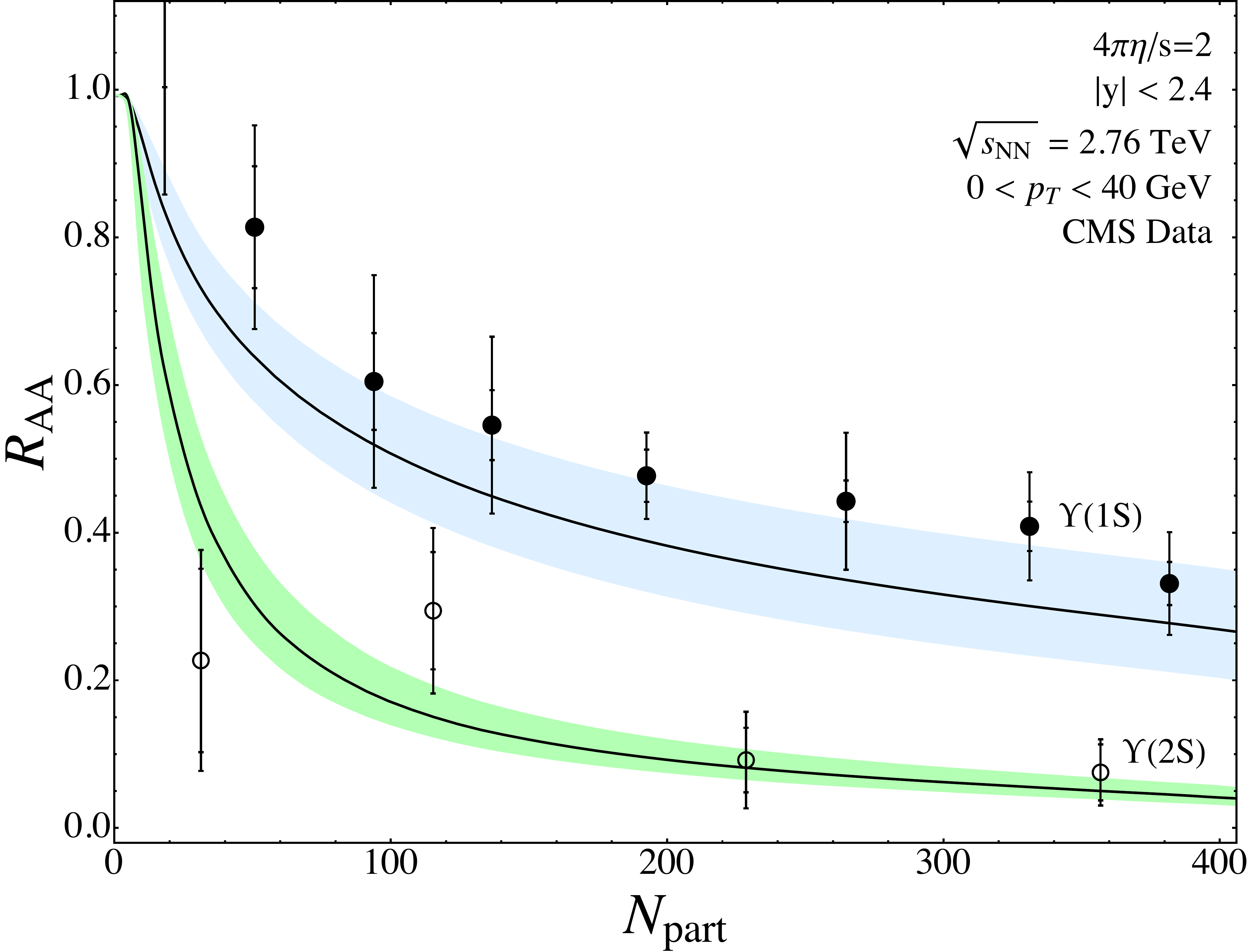}
\includegraphics[width=0.47\linewidth]{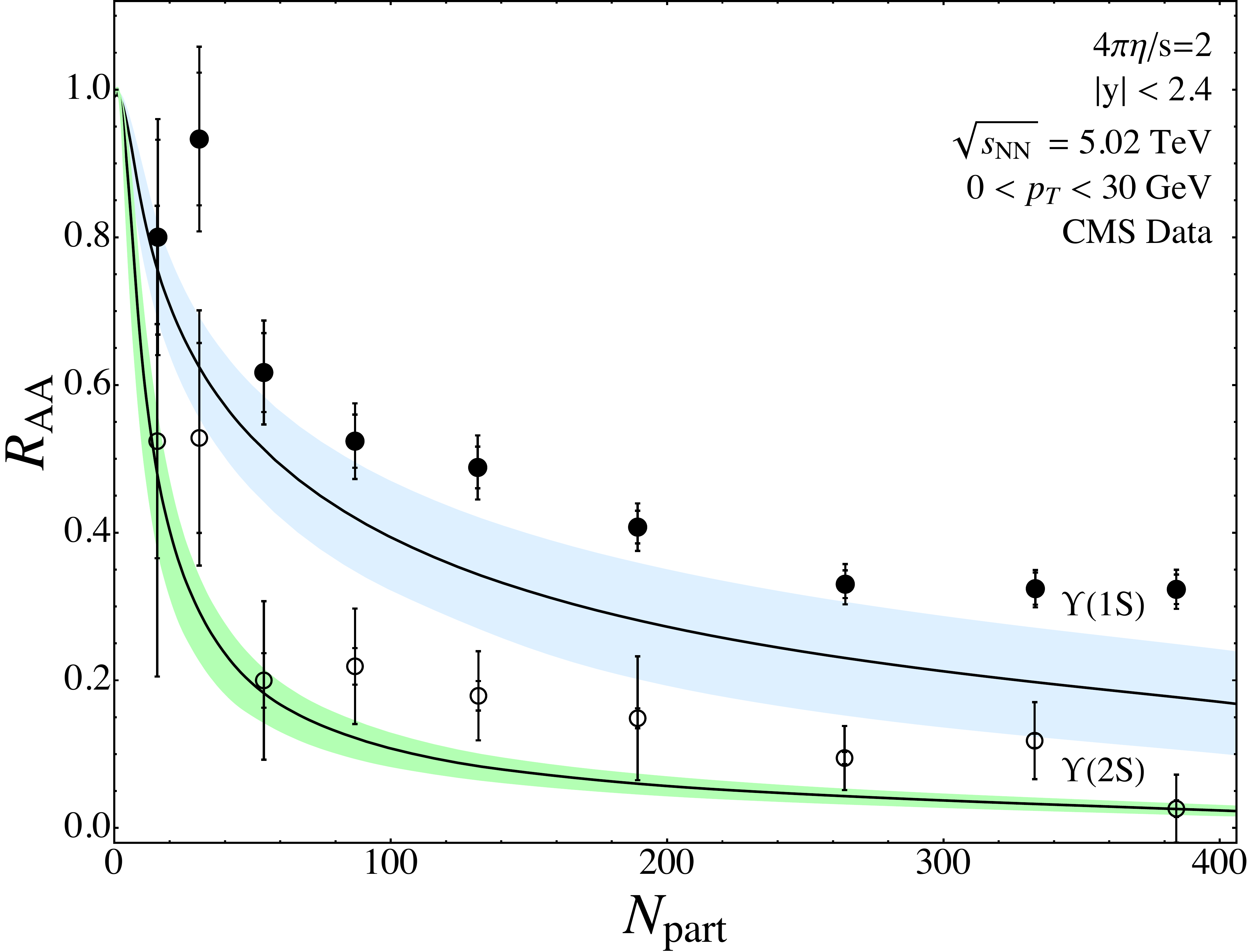}
}
\caption{
(Color online) $R_{AA}$ as a function of the number of participants $N_{\text{part}}$ compared to CMS data taken at the LHC for 2.76 TeV (left) and 5.02 TeV (right) Pb-Pb collisions. At 2.76 TeV (left) we find that within the uncertainty of the calculation we reproduce the ground state $R_{AA}$ with estimated having a slight tendency to take on lower values. The excited state $\Upsilon^\prime$ on the other hand is excellently captured. At 5.02 TeV (right) our estimates consistently lie below the experimental observations both for the ground state and the excited state, the deviation increasing with increasing centrality of the collisions.
}
\label{fig:RKNpart}
\end{figure*}

Turning to LHC energies, in Fig.~\ref{fig:RKNpart} we compare our new results with $R_{AA}$ data obtained by the CMS collaboration at (left) 2.76 TeV and (right) 5.02 TeV as function of centrality. Previous estimates of the $\Upsilon(1S)$ ground state suppression for LHC run1, based on the model potential, reproduced the data points best when selecting values of $4\pi\eta/s=2$ with stronger shear leading to systematically lower values \cite{Krouppa:2015yoa}. A similar conclusion was reached for the $\Upsilon(2S)$ suppression. Now with the lattice-vetted heavy quark potential the dependence on the assumed value of the shear viscosity is essentially absent and the stronger imaginary part in the lattice-vetted potential induces slightly stronger suppression. For $\Upsilon(1S)$ our new estimates agree with the data within the still relatively large error bars but are slightly lower than the experimental data. On the other hand, the trend in the excited state data points is excellently reproduced, touching also the point at the lowest centrality bin, providing a better description overall than the perturbative model results (see Ref.~\cite{Krouppa:2017lsw} for a compilation of the prior results).

That being said, when moving to the higher energy of run2 at the LHC, we find that the trend of stronger suppression continues in our estimates of bottomonium suppression. At this energy the lattice-vetted model overpredicts the amount of suppression of both the $\Upsilon(1S)$ and $\Upsilon(2S)$. This means that now our $R_{AA}$ systematically overestimates the suppression for both states. The discrepancy is larger for more central collisions while for smaller $N_{\rm part}$ we still find reasonable agreement with the data.

\begin{figure*}[h]
\centerline{
\includegraphics[width=0.47\linewidth]{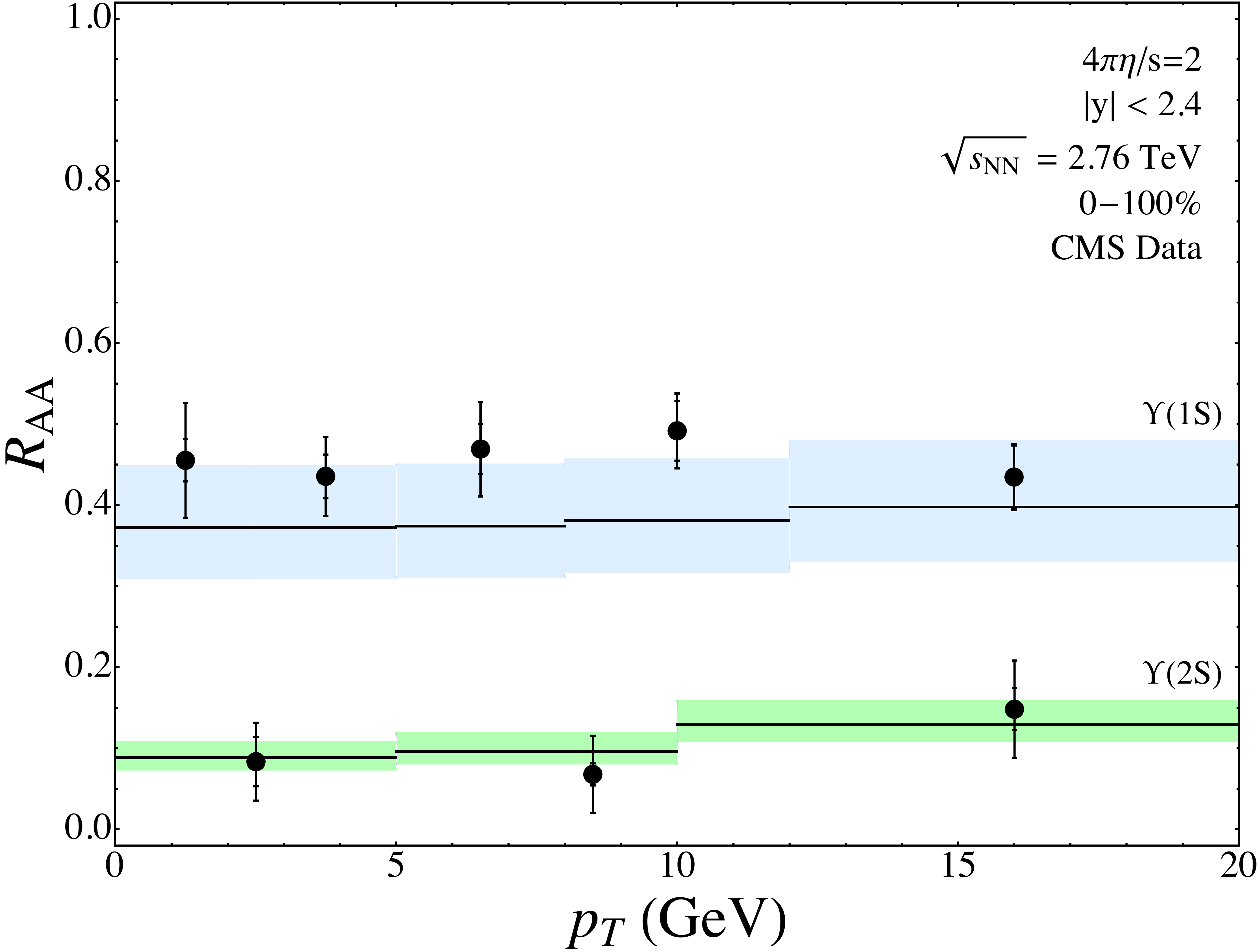}
\includegraphics[width=0.47\linewidth]{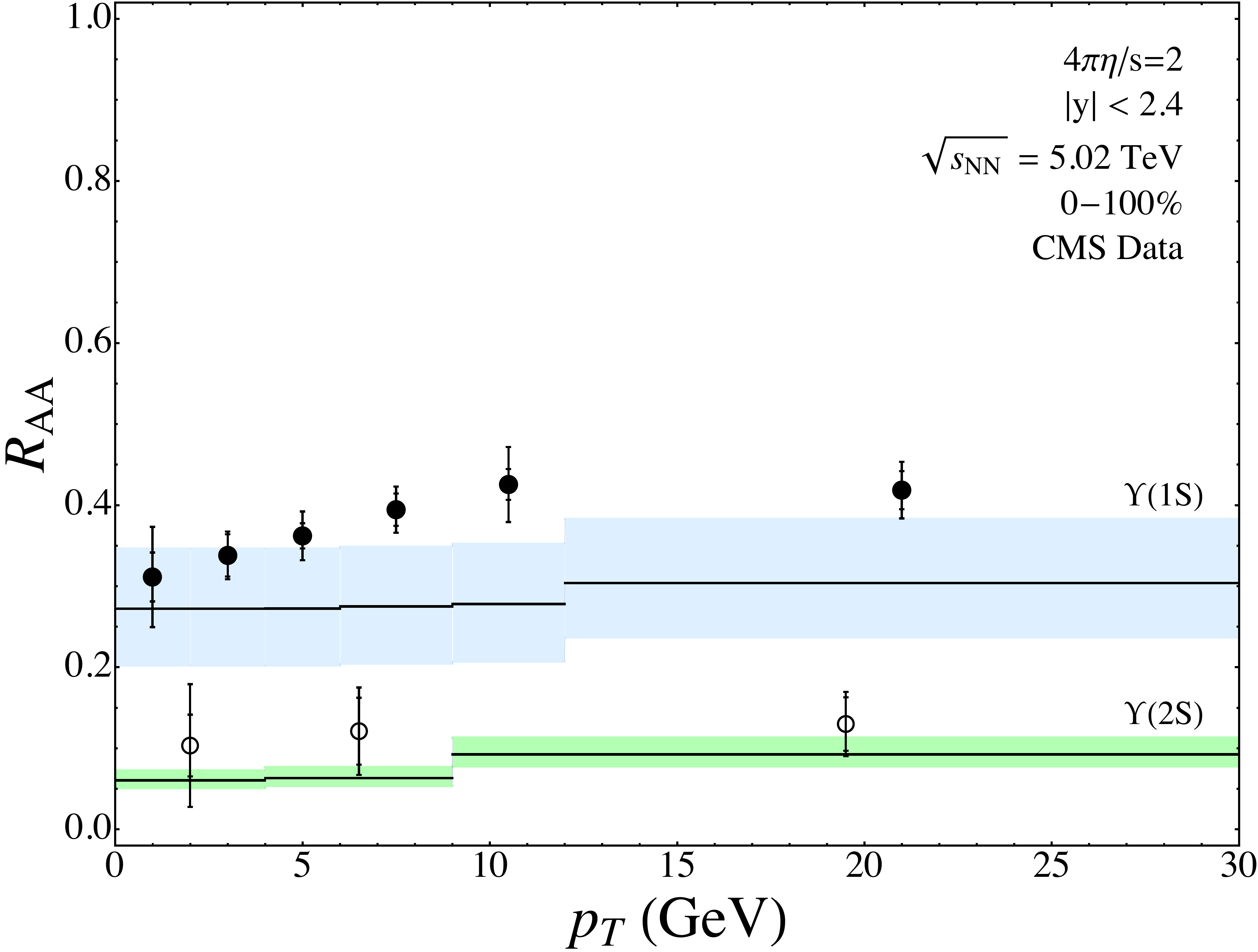}
}
\caption{
(Color online) $R_{AA}$ as a function of transverse momentum $p_{T}$ compared to CMS data taken at the LHC for 2.76 TeV (left) and 5.02 TeV (right) Pb-Pb collisions. Also here at LHC run1 energies we find good agreement with the $\Upsilon(1S)$ data and an excellent reproduction of the $\Upsilon(2S)$ suppression. At 5.02 TeV the experimentally determined suppression appears slightly weaker than what our calculation predicts, with good agreement for $\Upsilon(1S)$ at small $p_T$ and the largest discrepancies around 8-10GeV.}
\label{fig:RKpT}
\end{figure*}

In Fig. \ref{fig:RKpT} we plot the nuclear modification factor as function of $p_{T}$ integrated over all centrality classes at (left) 2.76 TeV and (right) 5.02 TeV. Similarly to our findings in terms of centrality at LHC run1 energies, the agreement here is best for $\Upsilon(2S)$ and acceptable for $\Upsilon(1S)$. There is a tendency visible to slightly overestimate the suppression when using the lattice-vetted potential in the computation.

\begin{figure*}[h]
\centerline{
\includegraphics[width=0.47\linewidth]{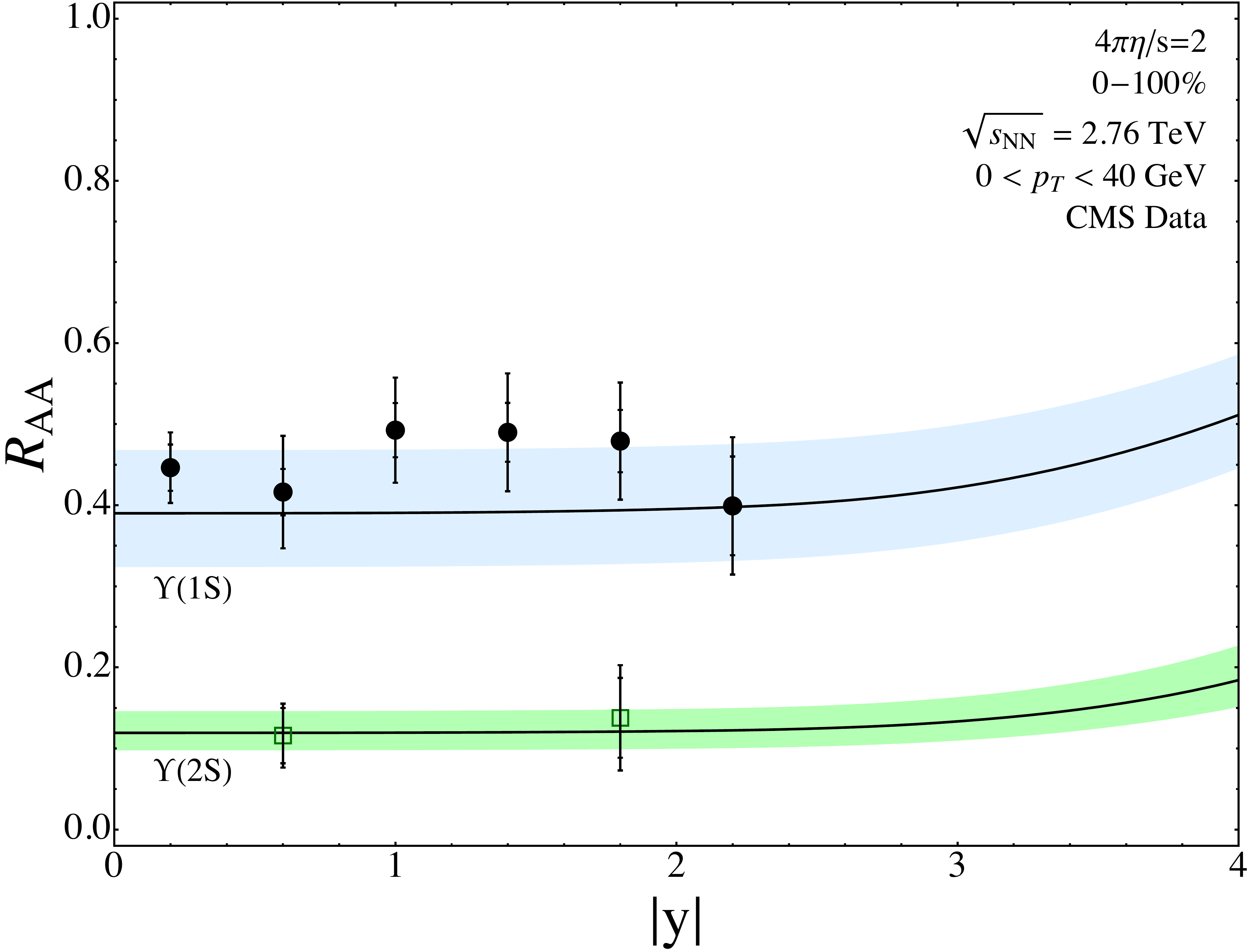}
\includegraphics[width=0.47\linewidth]{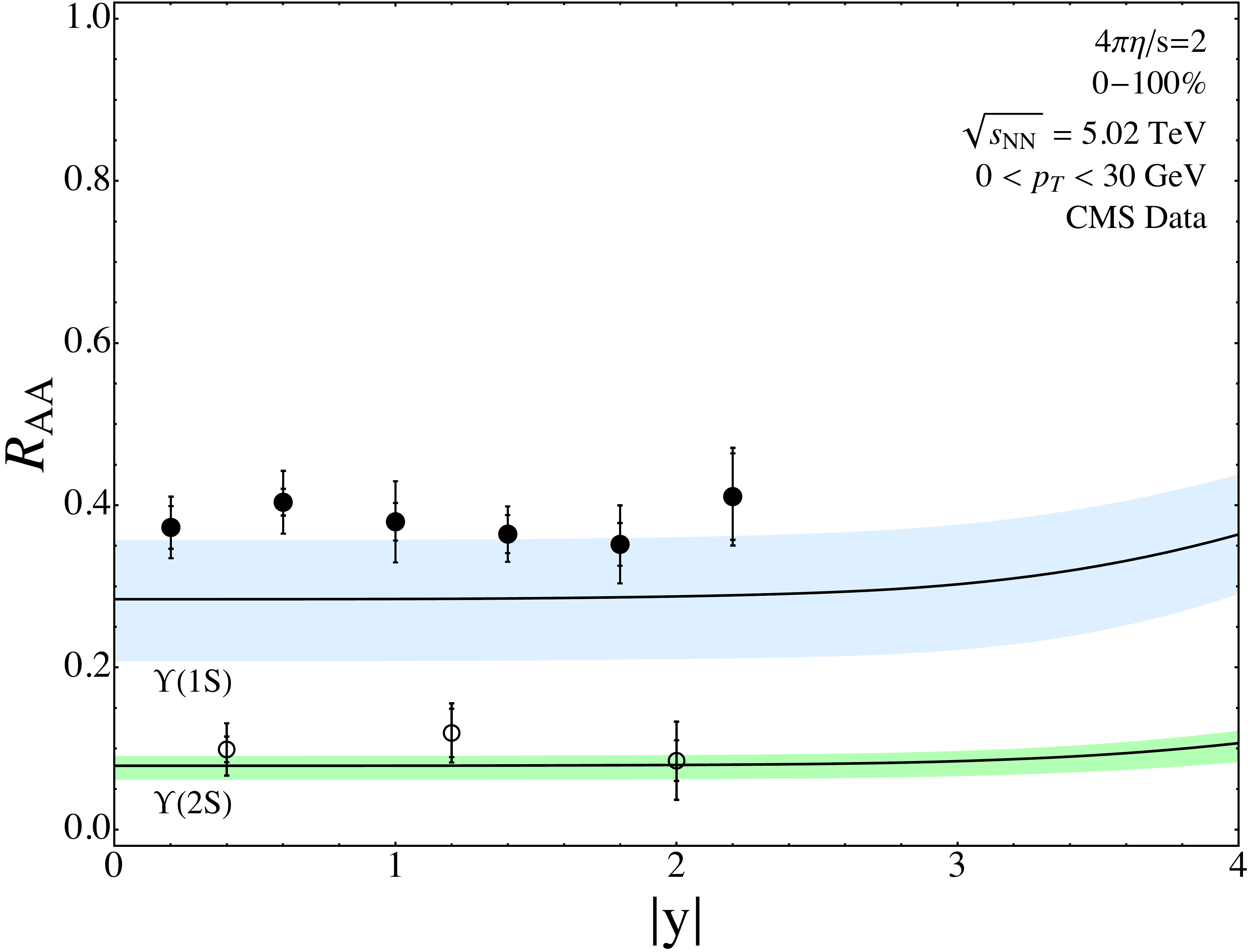}
}
\caption{
(Color online) $R_{AA}$ as a function of spatial rapidity $y$ compared to CMS data taken at the LHC for 2.76 TeV (left) and 5.02 TeV (right) Pb-Pb collisions. Similar to the $p_T$ plots, at LHC run1 energies we observe good agreement with the $\Upsilon(1S)$ data and the $\Upsilon(2S)$ suppression is very well reproduced. At 5.02 TeV the experimentally determined suppression appears slightly weaker than what our calculation predicts.
}
\label{fig:RKrap}
\end{figure*}

Finally, in Fig.~\ref{fig:RKrap} we show $R_{AA}$ as a function of spatial rapidity using the CMS cuts at 2.76 TeV (left) and 5.02 TeV (right) collisions at the LHC. The outcome of our calculation as can now be expected from the above discussion agrees well for the $\Upsilon(1S)$ at LHC run1 energies and gives an excellent account of the $\Upsilon(2S)$ suppression. At 5.02 TeV the trend of a overestimation of the suppression manifests itself again.


\section{Conclusions and outlook}
\label{sec:con}

By combining a lattice QCD vetted in-medium heavy-quark potential with a realistic anisotropic hydrodynamics simulation for the bulk matter created in a heavy-ion collision, we have studied the nuclear modification factor $R_{AA}$ for the bottomonium ground and first excited states. 
The functional form and temperature dependence of the lattice-vetted heavy quark potential lead to a significantly reduced dependence of the Upsilon $R_{AA}$ on the values of the shear to entropy ratio of the QGP. From a phenomenological point of view this is a welcome finding, since it strengthens the position of bottomonium as a genuine dynamical thermometer of the QGP, as originally envisioned.

We have found that our estimates for the suppression of primordial bottomonium show the best agreement with experimental results at RHIC energies, where they reproduce the measured data within errors. The larger imaginary part in the lattice-vetted potential compared to previously used model potentials leads to a stronger suppression, which in the case of RHIC energies is the reason for the excellent reproduction of the experimental data. At LHC run1 we find, on the other hand, hints of a tendency to overestimate the suppression compared to the data, while at 5.02 TeV we clearly see too much suppression when using the lattice-vetted potential. Two reasons for this behavior immediately come to mind. The first is related to the physics mechanism we assume underlies the imaginary part of the potential, the second one is related to the fact that no regeneration component has been included in our current study.

Our calculation uses a simple implementation of bottomonium suppression as discussed in Sec.~\ref{sec:dynamics}. By using the imaginary part of the potential directly in the Schr\"odinger equation we have assumed that it arises solely from gluo-dissociation of the heavy-quarkonium. Studies of bottomonium in perturbative pNRQCD have shown however that the imaginary part contains both contributions from gluo-dissociation and Landau damping, which leads to the excitation of bottomonium states without decay of the $Q\bar{Q}$ pair. Eventually we will need to understand how to disentangle these two contributions, since the latter, as shown in exploratory studies in the context of open-quantum systems  \cite{Rothkopf:2013kya,Kajimoto:2017rel} leads to a weaker suppression. I.e. our inability to disentangle the underlying mechanisms of the imaginary part have led us to assign it fully to a loss channel which may overestimate the actual suppression.

One possible way for future improvement of the proper potential based real-time description of the heavy quarkonium evolution, would be to utilize the framework of open-quantum systems. As was proposed e.g.~in \cite{Akamatsu:2011se}, the imaginary part of the heavy-quark potential, which arises in the description of the unequal time correlation functions of Eq.~\eqref{Eq:ForwProp} is unraveled into either a stochastic dynamics on the level of the Schr\"odinger equation or equivalently into a master equation for the density matrix of quarkonium states. Recent developments in this field include a formulation of a Lindblad type coupled dynamics for color singlet and octet degrees of freedom \cite{Akamatsu:2014qsa,Akamatsu:2015kaa,DeBoni:2017ocl}, which has been evaluated in a perturbative setting in \cite{Brambilla:2016wgg}. A first attempt to describe bottomonium dynamics was presented in \cite{Rothkopf:2013kya}, however, the medium evolution used in this paper was not yet realistic.

In addition, we have estimated the suppression of bottomonium states without including a possibility for regeneration. While for charmonium there are clear indications for the presence of a regeneration components in the observed yields \cite{Rapp:2017chc}, no such signals have been firmly established for bottomonium yet. One hint towards the onset of regeneration could be the small change in the bottomonium suppression when going from 2.76 TeV to 5.02 TeV. While the lattice-vetted potential predicts suppression that increases with beam energy, the data appears to not decrease as rapidly as suggested by our calculations.  This would allow for the addition of a regeneration component to approach the data from below, however, we are not yet in a position to quantitatively estimate the magnitude of this effect.

In the end it will be necessary to explore both paths to come to a robust phenomenological understanding of the observed yields. The model of bottomonium suppression needs to be made more flexible in order to accommodate the different underlying physics mechanisms inherent in the imaginary part, e.g. via a stochastic Sch\"odinger equation and the inclusion of regeneration via a rate equation framework is desirable.

\acknowledgements

A.R.~acknowledges fruitful discussions with Y. Akamatsu. A.R.~was supported by the DFG Collaborative Research Centre SFB
1225 (ISOQUANT).  B.K.~and~M.S. were supported by the U.S. Department of Energy, Office of Science, under Award No. DE-SC0013470.


\bibliography{bottom}

\begin{thebibliography}{84}%
\makeatletter
\providecommand \@ifxundefined [1]{%
 \@ifx{#1\undefined}
}%
\providecommand \@ifnum [1]{%
 \ifnum #1\expandafter \@firstoftwo
 \else \expandafter \@secondoftwo
 \fi
}%
\providecommand \@ifx [1]{%
 \ifx #1\expandafter \@firstoftwo
 \else \expandafter \@secondoftwo
 \fi
}%
\providecommand \natexlab [1]{#1}%
\providecommand \enquote  [1]{``#1''}%
\providecommand \bibnamefont  [1]{#1}%
\providecommand \bibfnamefont [1]{#1}%
\providecommand \citenamefont [1]{#1}%
\providecommand \href@noop [0]{\@secondoftwo}%
\providecommand \href [0]{\begingroup \@sanitize@url \@href}%
\providecommand \@href[1]{\@@startlink{#1}\@@href}%
\providecommand \@@href[1]{\endgroup#1\@@endlink}%
\providecommand \@sanitize@url [0]{\catcode `\\12\catcode `\$12\catcode
  `\&12\catcode `\#12\catcode `\^12\catcode `\_12\catcode `\%12\relax}%
\providecommand \@@startlink[1]{}%
\providecommand \@@endlink[0]{}%
\providecommand \url  [0]{\begingroup\@sanitize@url \@url }%
\providecommand \@url [1]{\endgroup\@href {#1}{\urlprefix }}%
\providecommand \urlprefix  [0]{URL }%
\providecommand \Eprint [0]{\href }%
\providecommand \doibase [0]{http://dx.doi.org/}%
\providecommand \selectlanguage [0]{\@gobble}%
\providecommand \bibinfo  [0]{\@secondoftwo}%
\providecommand \bibfield  [0]{\@secondoftwo}%
\providecommand \translation [1]{[#1]}%
\providecommand \BibitemOpen [0]{}%
\providecommand \bibitemStop [0]{}%
\providecommand \bibitemNoStop [0]{.\EOS\space}%
\providecommand \EOS [0]{\spacefactor3000\relax}%
\providecommand \BibitemShut  [1]{\csname bibitem#1\endcsname}%
\let\auto@bib@innerbib\@empty
\bibitem [{\citenamefont {Romatschke}(2010)}]{Romatschke:2009im}%
  \BibitemOpen
  \bibfield  {author} {\bibinfo {author} {\bibfnamefont {P.}~\bibnamefont
  {Romatschke}},\ }\href {\doibase 10.1142/S0218301310014613} {\bibfield
  {journal} {\bibinfo  {journal} {Int.J.Mod.Phys.}\ }\textbf {\bibinfo {volume}
  {E19}},\ \bibinfo {pages} {1} (\bibinfo {year} {2010})},\ \Eprint
  {http://arxiv.org/abs/0902.3663} {arXiv:0902.3663 [hep-ph]} \BibitemShut
  {NoStop}%
\bibitem [{\citenamefont {Heinz}\ and\ \citenamefont
  {Snellings}(2013)}]{Heinz:2013AnnRevNuc}%
  \BibitemOpen
  \bibfield  {author} {\bibinfo {author} {\bibfnamefont {U.}~\bibnamefont
  {Heinz}}\ and\ \bibinfo {author} {\bibfnamefont {R.}~\bibnamefont
  {Snellings}},\ }\href@noop {} {\bibfield  {journal} {\bibinfo  {journal}
  {Ann. Rev. Nucl. Part. Sci.}\ }\textbf {\bibinfo {volume} {63}} (\bibinfo
  {year} {2013})},\ \Eprint {http://arxiv.org/abs/1301.2826} {arXiv:1301.2826
  [nucl-th]} \BibitemShut {NoStop}%
\bibitem [{\citenamefont {Schenke}\ \emph {et~al.}(2011)\citenamefont
  {Schenke}, \citenamefont {Jeon},\ and\ \citenamefont
  {Gale}}]{Schenke:2011tv}%
  \BibitemOpen
  \bibfield  {author} {\bibinfo {author} {\bibfnamefont {B.}~\bibnamefont
  {Schenke}}, \bibinfo {author} {\bibfnamefont {S.}~\bibnamefont {Jeon}}, \
  and\ \bibinfo {author} {\bibfnamefont {C.}~\bibnamefont {Gale}},\ }\href
  {\doibase 10.1016/j.physletb.2011.06.065} {\bibfield  {journal} {\bibinfo
  {journal} {Phys.Lett.}\ }\textbf {\bibinfo {volume} {B702}},\ \bibinfo
  {pages} {59} (\bibinfo {year} {2011})},\ \Eprint
  {http://arxiv.org/abs/1102.0575} {arXiv:1102.0575 [hep-ph]} \BibitemShut
  {NoStop}%
\bibitem [{\citenamefont {Gale}\ \emph {et~al.}(2013)\citenamefont {Gale},
  \citenamefont {Jeon},\ and\ \citenamefont {Schenke}}]{Gale:2013IntJModPhys}%
  \BibitemOpen
  \bibfield  {author} {\bibinfo {author} {\bibfnamefont {C.}~\bibnamefont
  {Gale}}, \bibinfo {author} {\bibfnamefont {S.}~\bibnamefont {Jeon}}, \ and\
  \bibinfo {author} {\bibfnamefont {B.}~\bibnamefont {Schenke}},\ }\href@noop
  {} {\bibfield  {journal} {\bibinfo  {journal} {Int.J.Mod.Phys.}\ }\textbf
  {\bibinfo {volume} {A28}} (\bibinfo {year} {2013})},\ \Eprint
  {http://arxiv.org/abs/1301.5893} {arXiv:1301.5893 [nucl-th]} \BibitemShut
  {NoStop}%
\bibitem [{\citenamefont {Ryu}\ \emph {et~al.}(2015)\citenamefont {Ryu},
  \citenamefont {Paquet}, \citenamefont {Shen}, \citenamefont {Denicol},
  \citenamefont {Schenke}, \citenamefont {Jeon},\ and\ \citenamefont
  {Gale}}]{Ryu:2015vwa}%
  \BibitemOpen
  \bibfield  {author} {\bibinfo {author} {\bibfnamefont {S.}~\bibnamefont
  {Ryu}}, \bibinfo {author} {\bibfnamefont {J.~F.}\ \bibnamefont {Paquet}},
  \bibinfo {author} {\bibfnamefont {C.}~\bibnamefont {Shen}}, \bibinfo {author}
  {\bibfnamefont {G.~S.}\ \bibnamefont {Denicol}}, \bibinfo {author}
  {\bibfnamefont {B.}~\bibnamefont {Schenke}}, \bibinfo {author} {\bibfnamefont
  {S.}~\bibnamefont {Jeon}}, \ and\ \bibinfo {author} {\bibfnamefont
  {C.}~\bibnamefont {Gale}},\ }\href {\doibase 10.1103/PhysRevLett.115.132301}
  {\bibfield  {journal} {\bibinfo  {journal} {Phys. Rev. Lett.}\ }\textbf
  {\bibinfo {volume} {115}},\ \bibinfo {pages} {132301} (\bibinfo {year}
  {2015})},\ \Eprint {http://arxiv.org/abs/1502.01675} {arXiv:1502.01675
  [nucl-th]} \BibitemShut {NoStop}%
\bibitem [{\citenamefont {Niemi}\ \emph
  {et~al.}(2016{\natexlab{a}})\citenamefont {Niemi}, \citenamefont {Eskola},\
  and\ \citenamefont {Paatelainen}}]{Niemi:2015qia}%
  \BibitemOpen
  \bibfield  {author} {\bibinfo {author} {\bibfnamefont {H.}~\bibnamefont
  {Niemi}}, \bibinfo {author} {\bibfnamefont {K.~J.}\ \bibnamefont {Eskola}}, \
  and\ \bibinfo {author} {\bibfnamefont {R.}~\bibnamefont {Paatelainen}},\
  }\href {\doibase 10.1103/PhysRevC.93.024907} {\bibfield  {journal} {\bibinfo
  {journal} {Phys. Rev.}\ }\textbf {\bibinfo {volume} {C93}},\ \bibinfo {pages}
  {024907} (\bibinfo {year} {2016}{\natexlab{a}})},\ \Eprint
  {http://arxiv.org/abs/1505.02677} {arXiv:1505.02677 [hep-ph]} \BibitemShut
  {NoStop}%
\bibitem [{\citenamefont {Niemi}\ \emph
  {et~al.}(2016{\natexlab{b}})\citenamefont {Niemi}, \citenamefont {Eskola},
  \citenamefont {Paatelainen},\ and\ \citenamefont {Tuominen}}]{Niemi:2015voa}%
  \BibitemOpen
  \bibfield  {author} {\bibinfo {author} {\bibfnamefont {H.}~\bibnamefont
  {Niemi}}, \bibinfo {author} {\bibfnamefont {K.~J.}\ \bibnamefont {Eskola}},
  \bibinfo {author} {\bibfnamefont {R.}~\bibnamefont {Paatelainen}}, \ and\
  \bibinfo {author} {\bibfnamefont {K.}~\bibnamefont {Tuominen}},\ }\href
  {\doibase 10.1103/PhysRevC.93.014912} {\bibfield  {journal} {\bibinfo
  {journal} {Phys. Rev.}\ }\textbf {\bibinfo {volume} {C93}},\ \bibinfo {pages}
  {014912} (\bibinfo {year} {2016}{\natexlab{b}})},\ \Eprint
  {http://arxiv.org/abs/1511.04296} {arXiv:1511.04296 [hep-ph]} \BibitemShut
  {NoStop}%
\bibitem [{\citenamefont {Alqahtani}\ \emph
  {et~al.}(2017{\natexlab{a}})\citenamefont {Alqahtani}, \citenamefont
  {Nopoush}, \citenamefont {Ryblewski},\ and\ \citenamefont
  {Strickland}}]{Alqahtani:2017jwl}%
  \BibitemOpen
  \bibfield  {author} {\bibinfo {author} {\bibfnamefont {M.}~\bibnamefont
  {Alqahtani}}, \bibinfo {author} {\bibfnamefont {M.}~\bibnamefont {Nopoush}},
  \bibinfo {author} {\bibfnamefont {R.}~\bibnamefont {Ryblewski}}, \ and\
  \bibinfo {author} {\bibfnamefont {M.}~\bibnamefont {Strickland}},\ }\href
  {\doibase 10.1103/PhysRevLett.119.042301} {\bibfield  {journal} {\bibinfo
  {journal} {Phys. Rev. Lett.}\ }\textbf {\bibinfo {volume} {119}},\ \bibinfo
  {pages} {042301} (\bibinfo {year} {2017}{\natexlab{a}})},\ \Eprint
  {http://arxiv.org/abs/1703.05808} {arXiv:1703.05808 [nucl-th]} \BibitemShut
  {NoStop}%
\bibitem [{\citenamefont {Alqahtani}\ \emph
  {et~al.}(2017{\natexlab{b}})\citenamefont {Alqahtani}, \citenamefont
  {Nopoush}, \citenamefont {Ryblewski},\ and\ \citenamefont
  {Strickland}}]{Alqahtani:2017tnq}%
  \BibitemOpen
  \bibfield  {author} {\bibinfo {author} {\bibfnamefont {M.}~\bibnamefont
  {Alqahtani}}, \bibinfo {author} {\bibfnamefont {M.}~\bibnamefont {Nopoush}},
  \bibinfo {author} {\bibfnamefont {R.}~\bibnamefont {Ryblewski}}, \ and\
  \bibinfo {author} {\bibfnamefont {M.}~\bibnamefont {Strickland}},\
  }\href@noop {} {\  (\bibinfo {year} {2017}{\natexlab{b}})},\ \Eprint
  {http://arxiv.org/abs/1705.10191} {arXiv:1705.10191 [nucl-th]} \BibitemShut
  {NoStop}%
\bibitem [{\citenamefont {Matsui}\ and\ \citenamefont
  {Satz}(1986)}]{Matsui:1986dk}%
  \BibitemOpen
  \bibfield  {author} {\bibinfo {author} {\bibfnamefont {T.}~\bibnamefont
  {Matsui}}\ and\ \bibinfo {author} {\bibfnamefont {H.}~\bibnamefont {Satz}},\
  }\href {\doibase 10.1016/0370-2693(86)91404-8} {\bibfield  {journal}
  {\bibinfo  {journal} {Phys. Lett.}\ }\textbf {\bibinfo {volume} {B178}},\
  \bibinfo {pages} {416} (\bibinfo {year} {1986})}\BibitemShut {NoStop}%
\bibitem [{\citenamefont {Karsch}\ \emph {et~al.}(1988)\citenamefont {Karsch},
  \citenamefont {Mehr},\ and\ \citenamefont {Satz}}]{Karsch:1987pv}%
  \BibitemOpen
  \bibfield  {author} {\bibinfo {author} {\bibfnamefont {F.}~\bibnamefont
  {Karsch}}, \bibinfo {author} {\bibfnamefont {M.~T.}\ \bibnamefont {Mehr}}, \
  and\ \bibinfo {author} {\bibfnamefont {H.}~\bibnamefont {Satz}},\ }\href
  {\doibase 10.1007/BF01549722} {\bibfield  {journal} {\bibinfo  {journal} {Z.
  Phys.}\ }\textbf {\bibinfo {volume} {C37}},\ \bibinfo {pages} {617} (\bibinfo
  {year} {1988})}\BibitemShut {NoStop}%
\bibitem [{\citenamefont {Andronic}\ \emph {et~al.}(2016)\citenamefont
  {Andronic} \emph {et~al.}}]{Andronic:2015wma}%
  \BibitemOpen
  \bibfield  {author} {\bibinfo {author} {\bibfnamefont {A.}~\bibnamefont
  {Andronic}} \emph {et~al.},\ }\href {\doibase 10.1140/epjc/s10052-015-3819-5}
  {\bibfield  {journal} {\bibinfo  {journal} {Eur. Phys. J.}\ }\textbf
  {\bibinfo {volume} {C76}},\ \bibinfo {pages} {107} (\bibinfo {year}
  {2016})},\ \Eprint {http://arxiv.org/abs/1506.03981} {arXiv:1506.03981
  [nucl-ex]} \BibitemShut {NoStop}%
\bibitem [{\citenamefont {Mocsy}\ \emph {et~al.}(2013)\citenamefont {Mocsy},
  \citenamefont {Petreczky},\ and\ \citenamefont {Strickland}}]{Mocsy:2013syh}%
  \BibitemOpen
  \bibfield  {author} {\bibinfo {author} {\bibfnamefont {A.}~\bibnamefont
  {Mocsy}}, \bibinfo {author} {\bibfnamefont {P.}~\bibnamefont {Petreczky}}, \
  and\ \bibinfo {author} {\bibfnamefont {M.}~\bibnamefont {Strickland}},\
  }\href {\doibase 10.1142/S0217751X13400125} {\bibfield  {journal} {\bibinfo
  {journal} {Int.J.Mod.Phys.}\ }\textbf {\bibinfo {volume} {A28}},\ \bibinfo
  {pages} {1340012} (\bibinfo {year} {2013})},\ \Eprint
  {http://arxiv.org/abs/1302.2180} {arXiv:1302.2180 [hep-ph]} \BibitemShut
  {NoStop}%
\bibitem [{\citenamefont {Braun-Munzinger}\ and\ \citenamefont
  {Stachel}(2001)}]{BraunMunzinger:2000ep}%
  \BibitemOpen
  \bibfield  {author} {\bibinfo {author} {\bibfnamefont {P.}~\bibnamefont
  {Braun-Munzinger}}\ and\ \bibinfo {author} {\bibfnamefont {J.}~\bibnamefont
  {Stachel}},\ }\bibfield  {booktitle} {\emph {\bibinfo {booktitle} {{Nuclei
  and nucleons. Proceedings, International Symposium on the occasion of Achim
  Richter's 60th birthday, Darmstadt, Germany, October 11-13, 2000}}},\ }\href
  {\doibase 10.1016/S0375-9474(01)00936-8} {\bibfield  {journal} {\bibinfo
  {journal} {Nucl. Phys.}\ }\textbf {\bibinfo {volume} {A690}},\ \bibinfo
  {pages} {119} (\bibinfo {year} {2001})},\ \Eprint
  {http://arxiv.org/abs/nucl-th/0012064} {arXiv:nucl-th/0012064 [nucl-th]}
  \BibitemShut {NoStop}%
\bibitem [{\citenamefont {Thews}\ \emph {et~al.}(2001)\citenamefont {Thews},
  \citenamefont {Schroedter},\ and\ \citenamefont {Rafelski}}]{Thews:2000rj}%
  \BibitemOpen
  \bibfield  {author} {\bibinfo {author} {\bibfnamefont {R.~L.}\ \bibnamefont
  {Thews}}, \bibinfo {author} {\bibfnamefont {M.}~\bibnamefont {Schroedter}}, \
  and\ \bibinfo {author} {\bibfnamefont {J.}~\bibnamefont {Rafelski}},\ }\href
  {\doibase 10.1103/PhysRevC.63.054905} {\bibfield  {journal} {\bibinfo
  {journal} {Phys. Rev.}\ }\textbf {\bibinfo {volume} {C63}},\ \bibinfo {pages}
  {054905} (\bibinfo {year} {2001})},\ \Eprint
  {http://arxiv.org/abs/hep-ph/0007323} {arXiv:hep-ph/0007323 [hep-ph]}
  \BibitemShut {NoStop}%
\bibitem [{\citenamefont {Grandchamp}\ and\ \citenamefont
  {Rapp}(2002)}]{Grandchamp:2002wp}%
  \BibitemOpen
  \bibfield  {author} {\bibinfo {author} {\bibfnamefont {L.}~\bibnamefont
  {Grandchamp}}\ and\ \bibinfo {author} {\bibfnamefont {R.}~\bibnamefont
  {Rapp}},\ }\href {\doibase 10.1016/S0375-9474(02)01027-8} {\bibfield
  {journal} {\bibinfo  {journal} {Nucl. Phys.}\ }\textbf {\bibinfo {volume}
  {A709}},\ \bibinfo {pages} {415} (\bibinfo {year} {2002})},\ \Eprint
  {http://arxiv.org/abs/hep-ph/0205305} {arXiv:hep-ph/0205305 [hep-ph]}
  \BibitemShut {NoStop}%
\bibitem [{\citenamefont {Young}\ and\ \citenamefont
  {Shuryak}(2009)}]{Young:2008he}%
  \BibitemOpen
  \bibfield  {author} {\bibinfo {author} {\bibfnamefont {C.}~\bibnamefont
  {Young}}\ and\ \bibinfo {author} {\bibfnamefont {E.}~\bibnamefont
  {Shuryak}},\ }\href {\doibase 10.1103/PhysRevC.79.034907} {\bibfield
  {journal} {\bibinfo  {journal} {Phys. Rev.}\ }\textbf {\bibinfo {volume}
  {C79}},\ \bibinfo {pages} {034907} (\bibinfo {year} {2009})},\ \Eprint
  {http://arxiv.org/abs/0803.2866} {arXiv:0803.2866 [nucl-th]} \BibitemShut
  {NoStop}%
\bibitem [{\citenamefont {Young}\ and\ \citenamefont
  {Shuryak}(2010)}]{Young:2009tj}%
  \BibitemOpen
  \bibfield  {author} {\bibinfo {author} {\bibfnamefont {C.}~\bibnamefont
  {Young}}\ and\ \bibinfo {author} {\bibfnamefont {E.}~\bibnamefont
  {Shuryak}},\ }\href {\doibase 10.1103/PhysRevC.81.034905} {\bibfield
  {journal} {\bibinfo  {journal} {Phys. Rev.}\ }\textbf {\bibinfo {volume}
  {C81}},\ \bibinfo {pages} {034905} (\bibinfo {year} {2010})},\ \Eprint
  {http://arxiv.org/abs/0911.3080} {arXiv:0911.3080 [nucl-th]} \BibitemShut
  {NoStop}%
\bibitem [{\citenamefont {Emerick}\ \emph {et~al.}(2012)\citenamefont
  {Emerick}, \citenamefont {Zhao},\ and\ \citenamefont
  {Rapp}}]{Emerick:2011xu}%
  \BibitemOpen
  \bibfield  {author} {\bibinfo {author} {\bibfnamefont {A.}~\bibnamefont
  {Emerick}}, \bibinfo {author} {\bibfnamefont {X.}~\bibnamefont {Zhao}}, \
  and\ \bibinfo {author} {\bibfnamefont {R.}~\bibnamefont {Rapp}},\ }\href
  {\doibase 10.1140/epja/i2012-12072-y} {\bibfield  {journal} {\bibinfo
  {journal} {Eur. Phys. J.}\ }\textbf {\bibinfo {volume} {A48}},\ \bibinfo
  {pages} {72} (\bibinfo {year} {2012})},\ \Eprint
  {http://arxiv.org/abs/1111.6537} {arXiv:1111.6537 [hep-ph]} \BibitemShut
  {NoStop}%
\bibitem [{\citenamefont {Rapp}\ and\ \citenamefont {Du}(2017)}]{Rapp:2017chc}%
  \BibitemOpen
  \bibfield  {author} {\bibinfo {author} {\bibfnamefont {R.}~\bibnamefont
  {Rapp}}\ and\ \bibinfo {author} {\bibfnamefont {X.}~\bibnamefont {Du}},\ }in\
  \href {http://inspirehep.net/record/1596910/files/arXiv:1704.07923.pdf}
  {\emph {\bibinfo {booktitle} {{26th International Conference on
  Ultrarelativistic Nucleus-Nucleus Collisions (Quark Matter 2017)
  Chicago,Illinois, USA, February 6-11, 2017}}}}\ (\bibinfo {year} {2017})\
  \Eprint {http://arxiv.org/abs/1704.07923} {arXiv:1704.07923 [hep-ph]}
  \BibitemShut {NoStop}%
\bibitem [{\citenamefont {Du}\ \emph {et~al.}(2017{\natexlab{a}})\citenamefont
  {Du}, \citenamefont {Rapp},\ and\ \citenamefont {He}}]{Du:2017qkv}%
  \BibitemOpen
  \bibfield  {author} {\bibinfo {author} {\bibfnamefont {X.}~\bibnamefont
  {Du}}, \bibinfo {author} {\bibfnamefont {R.}~\bibnamefont {Rapp}}, \ and\
  \bibinfo {author} {\bibfnamefont {M.}~\bibnamefont {He}},\ }\href@noop {} {\
  (\bibinfo {year} {2017}{\natexlab{a}})},\ \Eprint
  {http://arxiv.org/abs/1706.08670} {arXiv:1706.08670 [hep-ph]} \BibitemShut
  {NoStop}%
\bibitem [{\citenamefont {Abbas}\ \emph {et~al.}(2013)\citenamefont {Abbas}
  \emph {et~al.}}]{ALICE:2013xna}%
  \BibitemOpen
  \bibfield  {author} {\bibinfo {author} {\bibfnamefont {E.}~\bibnamefont
  {Abbas}} \emph {et~al.} (\bibinfo {collaboration} {ALICE}),\ }\href {\doibase
  10.1103/PhysRevLett.111.162301} {\bibfield  {journal} {\bibinfo  {journal}
  {Phys. Rev. Lett.}\ }\textbf {\bibinfo {volume} {111}},\ \bibinfo {pages}
  {162301} (\bibinfo {year} {2013})},\ \Eprint {http://arxiv.org/abs/1303.5880}
  {arXiv:1303.5880 [nucl-ex]} \BibitemShut {NoStop}%
\bibitem [{\citenamefont {Liu}\ \emph {et~al.}(2010)\citenamefont {Liu},
  \citenamefont {Xu},\ and\ \citenamefont {Zhuang}}]{Liu:2009gx}%
  \BibitemOpen
  \bibfield  {author} {\bibinfo {author} {\bibfnamefont {Y.}~\bibnamefont
  {Liu}}, \bibinfo {author} {\bibfnamefont {N.}~\bibnamefont {Xu}}, \ and\
  \bibinfo {author} {\bibfnamefont {P.}~\bibnamefont {Zhuang}},\ }\bibfield
  {booktitle} {\emph {\bibinfo {booktitle} {{Nucleus nucleus collisions.
  Proceedings, 10th International Conference, NN2009, Beijing, P.R. China,
  August 16-21, 2009}}},\ }\href {\doibase 10.1016/j.nuclphysa.2010.01.008}
  {\bibfield  {journal} {\bibinfo  {journal} {Nucl. Phys.}\ }\textbf {\bibinfo
  {volume} {A834}},\ \bibinfo {pages} {317C} (\bibinfo {year} {2010})},\
  \Eprint {http://arxiv.org/abs/0910.0959} {arXiv:0910.0959 [nucl-th]}
  \BibitemShut {NoStop}%
\bibitem [{\citenamefont {Zhao}\ \emph {et~al.}(2013)\citenamefont {Zhao},
  \citenamefont {Emerick},\ and\ \citenamefont {Rapp}}]{Zhao:2012gc}%
  \BibitemOpen
  \bibfield  {author} {\bibinfo {author} {\bibfnamefont {X.}~\bibnamefont
  {Zhao}}, \bibinfo {author} {\bibfnamefont {A.}~\bibnamefont {Emerick}}, \
  and\ \bibinfo {author} {\bibfnamefont {R.}~\bibnamefont {Rapp}},\ }\bibfield
  {booktitle} {\emph {\bibinfo {booktitle} {{Proceedings, 23rd International
  Conference on Ultrarelativistic Nucleus-Nucleus Collisions : Quark Matter
  2012 (QM 2012): Washington, DC, USA, August 13-18, 2012}}},\ }\href {\doibase
  10.1016/j.nuclphysa.2013.02.088} {\bibfield  {journal} {\bibinfo  {journal}
  {Nucl. Phys.}\ }\textbf {\bibinfo {volume} {A904-905}},\ \bibinfo {pages}
  {611c} (\bibinfo {year} {2013})},\ \Eprint {http://arxiv.org/abs/1210.6583}
  {arXiv:1210.6583 [hep-ph]} \BibitemShut {NoStop}%
\bibitem [{\citenamefont {Ryblewski}(2013)}]{Ryblewski:2013jsa}%
  \BibitemOpen
  \bibfield  {author} {\bibinfo {author} {\bibfnamefont {R.}~\bibnamefont
  {Ryblewski}},\ }\href {\doibase 10.1088/0954-3899/40/9/093101} {\bibfield
  {journal} {\bibinfo  {journal} {J.Phys.}\ }\textbf {\bibinfo {volume}
  {G40}},\ \bibinfo {pages} {093101} (\bibinfo {year} {2013})}\BibitemShut
  {NoStop}%
\bibitem [{\citenamefont {Strickland}(2014)}]{Strickland:2014pga}%
  \BibitemOpen
  \bibfield  {author} {\bibinfo {author} {\bibfnamefont {M.}~\bibnamefont
  {Strickland}},\ }\bibfield  {booktitle} {\emph {\bibinfo {booktitle} {{54th
  Cracow School of Theoretical Physics: QCD meets experiment}}},\ }\href
  {\doibase 10.5506/APhysPolB.45.2355} {\bibfield  {journal} {\bibinfo
  {journal} {Acta Phys. Polon.}\ }\textbf {\bibinfo {volume} {B45}},\ \bibinfo
  {pages} {2355} (\bibinfo {year} {2014})},\ \Eprint
  {http://arxiv.org/abs/1410.5786} {arXiv:1410.5786 [nucl-th]} \BibitemShut
  {NoStop}%
\bibitem [{\citenamefont {Dumitru}\ \emph {et~al.}(2008)\citenamefont
  {Dumitru}, \citenamefont {Guo},\ and\ \citenamefont
  {Strickland}}]{Dumitru:2007hy}%
  \BibitemOpen
  \bibfield  {author} {\bibinfo {author} {\bibfnamefont {A.}~\bibnamefont
  {Dumitru}}, \bibinfo {author} {\bibfnamefont {Y.}~\bibnamefont {Guo}}, \ and\
  \bibinfo {author} {\bibfnamefont {M.}~\bibnamefont {Strickland}},\ }\href
  {\doibase 10.1016/j.physletb.2008.02.048} {\bibfield  {journal} {\bibinfo
  {journal} {Phys.Lett.}\ }\textbf {\bibinfo {volume} {B662}},\ \bibinfo
  {pages} {37} (\bibinfo {year} {2008})},\ \Eprint
  {http://arxiv.org/abs/0711.4722} {arXiv:0711.4722 [hep-ph]} \BibitemShut
  {NoStop}%
\bibitem [{\citenamefont {Burnier}\ \emph {et~al.}(2009)\citenamefont
  {Burnier}, \citenamefont {Laine},\ and\ \citenamefont
  {Vepsalainen}}]{Burnier:2009yu}%
  \BibitemOpen
  \bibfield  {author} {\bibinfo {author} {\bibfnamefont {Y.}~\bibnamefont
  {Burnier}}, \bibinfo {author} {\bibfnamefont {M.}~\bibnamefont {Laine}}, \
  and\ \bibinfo {author} {\bibfnamefont {M.}~\bibnamefont {Vepsalainen}},\
  }\href {\doibase 10.1016/j.physletb.2009.05.067} {\bibfield  {journal}
  {\bibinfo  {journal} {Phys.Lett.}\ }\textbf {\bibinfo {volume} {B678}},\
  \bibinfo {pages} {86} (\bibinfo {year} {2009})},\ \Eprint
  {http://arxiv.org/abs/0903.3467} {arXiv:0903.3467 [hep-ph]} \BibitemShut
  {NoStop}%
\bibitem [{\citenamefont {Dumitru}\ \emph {et~al.}(2009)\citenamefont
  {Dumitru}, \citenamefont {Guo},\ and\ \citenamefont
  {Strickland}}]{Dumitru:2009fy}%
  \BibitemOpen
  \bibfield  {author} {\bibinfo {author} {\bibfnamefont {A.}~\bibnamefont
  {Dumitru}}, \bibinfo {author} {\bibfnamefont {Y.}~\bibnamefont {Guo}}, \ and\
  \bibinfo {author} {\bibfnamefont {M.}~\bibnamefont {Strickland}},\ }\href
  {\doibase 10.1103/PhysRevD.79.114003} {\bibfield  {journal} {\bibinfo
  {journal} {Phys.Rev.}\ }\textbf {\bibinfo {volume} {D79}},\ \bibinfo {pages}
  {114003} (\bibinfo {year} {2009})},\ \Eprint {http://arxiv.org/abs/0903.4703}
  {arXiv:0903.4703 [hep-ph]} \BibitemShut {NoStop}%
\bibitem [{\citenamefont {Strickland}(2011)}]{Strickland:2011mw}%
  \BibitemOpen
  \bibfield  {author} {\bibinfo {author} {\bibfnamefont {M.}~\bibnamefont
  {Strickland}},\ }\href {\doibase 10.1103/PhysRevLett.107.132301} {\bibfield
  {journal} {\bibinfo  {journal} {Phys.Rev.Lett.}\ }\textbf {\bibinfo {volume}
  {107}},\ \bibinfo {pages} {132301} (\bibinfo {year} {2011})},\ \Eprint
  {http://arxiv.org/abs/1106.2571} {arXiv:1106.2571 [hep-ph]} \BibitemShut
  {NoStop}%
\bibitem [{\citenamefont {Strickland}\ and\ \citenamefont
  {Bazow}(2012)}]{Strickland:2011aa}%
  \BibitemOpen
  \bibfield  {author} {\bibinfo {author} {\bibfnamefont {M.}~\bibnamefont
  {Strickland}}\ and\ \bibinfo {author} {\bibfnamefont {D.}~\bibnamefont
  {Bazow}},\ }\href {\doibase 10.1016/j.nuclphysa.2012.02.003} {\bibfield
  {journal} {\bibinfo  {journal} {Nucl. Phys.}\ }\textbf {\bibinfo {volume}
  {A879}},\ \bibinfo {pages} {25} (\bibinfo {year} {2012})},\ \Eprint
  {http://arxiv.org/abs/1112.2761} {arXiv:1112.2761 [nucl-th]} \BibitemShut
  {NoStop}%
\bibitem [{\citenamefont {Krouppa}\ \emph {et~al.}(2015)\citenamefont
  {Krouppa}, \citenamefont {Ryblewski},\ and\ \citenamefont
  {Strickland}}]{Krouppa:2015yoa}%
  \BibitemOpen
  \bibfield  {author} {\bibinfo {author} {\bibfnamefont {B.}~\bibnamefont
  {Krouppa}}, \bibinfo {author} {\bibfnamefont {R.}~\bibnamefont {Ryblewski}},
  \ and\ \bibinfo {author} {\bibfnamefont {M.}~\bibnamefont {Strickland}},\
  }\href {\doibase 10.1103/PhysRevC.92.061901} {\bibfield  {journal} {\bibinfo
  {journal} {Phys. Rev.}\ }\textbf {\bibinfo {volume} {C92}},\ \bibinfo {pages}
  {061901} (\bibinfo {year} {2015})},\ \Eprint
  {http://arxiv.org/abs/1507.03951} {arXiv:1507.03951 [hep-ph]} \BibitemShut
  {NoStop}%
\bibitem [{\citenamefont {Du}\ \emph {et~al.}(2017{\natexlab{b}})\citenamefont
  {Du}, \citenamefont {Dumitru}, \citenamefont {Guo},\ and\ \citenamefont
  {Strickland}}]{Du:2016wdx}%
  \BibitemOpen
  \bibfield  {author} {\bibinfo {author} {\bibfnamefont {Q.}~\bibnamefont
  {Du}}, \bibinfo {author} {\bibfnamefont {A.}~\bibnamefont {Dumitru}},
  \bibinfo {author} {\bibfnamefont {Y.}~\bibnamefont {Guo}}, \ and\ \bibinfo
  {author} {\bibfnamefont {M.}~\bibnamefont {Strickland}},\ }\href {\doibase
  10.1007/JHEP01(2017)123} {\bibfield  {journal} {\bibinfo  {journal} {JHEP}\
  }\textbf {\bibinfo {volume} {01}},\ \bibinfo {pages} {123} (\bibinfo {year}
  {2017}{\natexlab{b}})},\ \Eprint {http://arxiv.org/abs/1611.08379}
  {arXiv:1611.08379 [hep-ph]} \BibitemShut {NoStop}%
\bibitem [{\citenamefont {Krouppa}\ and\ \citenamefont
  {Strickland}(2016)}]{Krouppa:2016jcl}%
  \BibitemOpen
  \bibfield  {author} {\bibinfo {author} {\bibfnamefont {B.}~\bibnamefont
  {Krouppa}}\ and\ \bibinfo {author} {\bibfnamefont {M.}~\bibnamefont
  {Strickland}},\ }\href {\doibase 10.3390/universe2030016} {\bibfield
  {journal} {\bibinfo  {journal} {Universe}\ }\textbf {\bibinfo {volume} {2}},\
  \bibinfo {pages} {16} (\bibinfo {year} {2016})},\ \Eprint
  {http://arxiv.org/abs/1605.03561} {arXiv:1605.03561 [hep-ph]} \BibitemShut
  {NoStop}%
\bibitem [{\citenamefont {Biondini}\ \emph {et~al.}(2017)\citenamefont
  {Biondini}, \citenamefont {Brambilla}, \citenamefont {Escobedo},\ and\
  \citenamefont {Vairo}}]{Biondini:2017qjh}%
  \BibitemOpen
  \bibfield  {author} {\bibinfo {author} {\bibfnamefont {S.}~\bibnamefont
  {Biondini}}, \bibinfo {author} {\bibfnamefont {N.}~\bibnamefont {Brambilla}},
  \bibinfo {author} {\bibfnamefont {M.~A.}\ \bibnamefont {Escobedo}}, \ and\
  \bibinfo {author} {\bibfnamefont {A.}~\bibnamefont {Vairo}},\ }\href
  {\doibase 10.1103/PhysRevD.95.074016} {\bibfield  {journal} {\bibinfo
  {journal} {Phys. Rev.}\ }\textbf {\bibinfo {volume} {D95}},\ \bibinfo {pages}
  {074016} (\bibinfo {year} {2017})},\ \Eprint
  {http://arxiv.org/abs/1701.06956} {arXiv:1701.06956 [hep-ph]} \BibitemShut
  {NoStop}%
\bibitem [{\citenamefont {Krouppa}\ \emph {et~al.}(2017)\citenamefont
  {Krouppa}, \citenamefont {Ryblewski},\ and\ \citenamefont
  {Strickland}}]{Krouppa:2017lsw}%
  \BibitemOpen
  \bibfield  {author} {\bibinfo {author} {\bibfnamefont {B.}~\bibnamefont
  {Krouppa}}, \bibinfo {author} {\bibfnamefont {R.}~\bibnamefont {Ryblewski}},
  \ and\ \bibinfo {author} {\bibfnamefont {M.}~\bibnamefont {Strickland}},\
  }in\ \href {https://inspirehep.net/record/1590902/files/arXiv:1704.02361.pdf}
  {\emph {\bibinfo {booktitle} {{26th International Conference on
  Ultrarelativistic Nucleus-Nucleus Collisions (Quark Matter 2017)
  Chicago,Illinois, USA, February 6-11, 2017}}}}\ (\bibinfo {year} {2017})\
  \Eprint {http://arxiv.org/abs/1704.02361} {arXiv:1704.02361 [nucl-th]}
  \BibitemShut {NoStop}%
\bibitem [{\citenamefont {Nopoush}\ \emph {et~al.}(2017)\citenamefont
  {Nopoush}, \citenamefont {Guo},\ and\ \citenamefont
  {Strickland}}]{Nopoush:2017zbu}%
  \BibitemOpen
  \bibfield  {author} {\bibinfo {author} {\bibfnamefont {M.}~\bibnamefont
  {Nopoush}}, \bibinfo {author} {\bibfnamefont {Y.}~\bibnamefont {Guo}}, \ and\
  \bibinfo {author} {\bibfnamefont {M.}~\bibnamefont {Strickland}},\ }\href
  {\doibase 10.1007/JHEP09(2017)063} {\bibfield  {journal} {\bibinfo  {journal}
  {JHEP}\ }\textbf {\bibinfo {volume} {09}},\ \bibinfo {pages} {063} (\bibinfo
  {year} {2017})},\ \Eprint {http://arxiv.org/abs/1706.08091} {arXiv:1706.08091
  [hep-ph]} \BibitemShut {NoStop}%
\bibitem [{\citenamefont {Romatschke}\ and\ \citenamefont
  {Strickland}(2003)}]{Romatschke:2003ms}%
  \BibitemOpen
  \bibfield  {author} {\bibinfo {author} {\bibfnamefont {P.}~\bibnamefont
  {Romatschke}}\ and\ \bibinfo {author} {\bibfnamefont {M.}~\bibnamefont
  {Strickland}},\ }\href@noop {} {\bibfield  {journal} {\bibinfo  {journal}
  {Phys. Rev.}\ }\textbf {\bibinfo {volume} {D68}},\ \bibinfo {pages} {036004}
  (\bibinfo {year} {2003})},\ \Eprint {http://arxiv.org/abs/hep-ph/0304092}
  {hep-ph/0304092} \BibitemShut {NoStop}%
\bibitem [{\citenamefont {Martinez}\ and\ \citenamefont
  {Strickland}(2010)}]{Martinez:2010sc}%
  \BibitemOpen
  \bibfield  {author} {\bibinfo {author} {\bibfnamefont {M.}~\bibnamefont
  {Martinez}}\ and\ \bibinfo {author} {\bibfnamefont {M.}~\bibnamefont
  {Strickland}},\ }\href {\doibase 10.1016/j.nuclphysa.2010.08.011} {\bibfield
  {journal} {\bibinfo  {journal} {Nucl. Phys.}\ }\textbf {\bibinfo {volume}
  {A848}},\ \bibinfo {pages} {183} (\bibinfo {year} {2010})},\ \Eprint
  {http://arxiv.org/abs/1007.0889} {arXiv:1007.0889 [nucl-th]} \BibitemShut
  {NoStop}%
\bibitem [{\citenamefont {Florkowski}\ and\ \citenamefont
  {Ryblewski}(2011)}]{Florkowski:2010cf}%
  \BibitemOpen
  \bibfield  {author} {\bibinfo {author} {\bibfnamefont {W.}~\bibnamefont
  {Florkowski}}\ and\ \bibinfo {author} {\bibfnamefont {R.}~\bibnamefont
  {Ryblewski}},\ }\href {\doibase 10.1103/PhysRevC.83.034907} {\bibfield
  {journal} {\bibinfo  {journal} {Phys.Rev.}\ }\textbf {\bibinfo {volume}
  {C83}},\ \bibinfo {pages} {034907} (\bibinfo {year} {2011})},\ \Eprint
  {http://arxiv.org/abs/1007.0130} {arXiv:1007.0130 [nucl-th]} \BibitemShut
  {NoStop}%
\bibitem [{\citenamefont {Nopoush}\ \emph
  {et~al.}(2014{\natexlab{a}})\citenamefont {Nopoush}, \citenamefont
  {Ryblewski},\ and\ \citenamefont {Strickland}}]{Nopoush:2014pfa}%
  \BibitemOpen
  \bibfield  {author} {\bibinfo {author} {\bibfnamefont {M.}~\bibnamefont
  {Nopoush}}, \bibinfo {author} {\bibfnamefont {R.}~\bibnamefont {Ryblewski}},
  \ and\ \bibinfo {author} {\bibfnamefont {M.}~\bibnamefont {Strickland}},\
  }\href {\doibase 10.1103/PhysRevC.90.014908} {\bibfield  {journal} {\bibinfo
  {journal} {Phys.Rev.}\ }\textbf {\bibinfo {volume} {C90}},\ \bibinfo {pages}
  {014908} (\bibinfo {year} {2014}{\natexlab{a}})},\ \Eprint
  {http://arxiv.org/abs/1405.1355} {arXiv:1405.1355 [hep-ph]} \BibitemShut
  {NoStop}%
\bibitem [{\citenamefont {Alqahtani}\ \emph {et~al.}(2015)\citenamefont
  {Alqahtani}, \citenamefont {Nopoush},\ and\ \citenamefont
  {Strickland}}]{Alqahtani:2015qja}%
  \BibitemOpen
  \bibfield  {author} {\bibinfo {author} {\bibfnamefont {M.}~\bibnamefont
  {Alqahtani}}, \bibinfo {author} {\bibfnamefont {M.}~\bibnamefont {Nopoush}},
  \ and\ \bibinfo {author} {\bibfnamefont {M.}~\bibnamefont {Strickland}},\
  }\href {\doibase 10.1103/PhysRevC.92.054910} {\bibfield  {journal} {\bibinfo
  {journal} {Phys. Rev.}\ }\textbf {\bibinfo {volume} {C92}},\ \bibinfo {pages}
  {054910} (\bibinfo {year} {2015})},\ \Eprint
  {http://arxiv.org/abs/1509.02913} {arXiv:1509.02913 [hep-ph]} \BibitemShut
  {NoStop}%
\bibitem [{\citenamefont {Alqahtani}\ \emph
  {et~al.}(2017{\natexlab{c}})\citenamefont {Alqahtani}, \citenamefont
  {Nopoush},\ and\ \citenamefont {Strickland}}]{Alqahtani:2016rth}%
  \BibitemOpen
  \bibfield  {author} {\bibinfo {author} {\bibfnamefont {M.}~\bibnamefont
  {Alqahtani}}, \bibinfo {author} {\bibfnamefont {M.}~\bibnamefont {Nopoush}},
  \ and\ \bibinfo {author} {\bibfnamefont {M.}~\bibnamefont {Strickland}},\
  }\href {\doibase 10.1103/PhysRevC.95.034906} {\bibfield  {journal} {\bibinfo
  {journal} {Phys. Rev.}\ }\textbf {\bibinfo {volume} {C95}},\ \bibinfo {pages}
  {034906} (\bibinfo {year} {2017}{\natexlab{c}})},\ \Eprint
  {http://arxiv.org/abs/1605.02101} {arXiv:1605.02101 [nucl-th]} \BibitemShut
  {NoStop}%
\bibitem [{\citenamefont {Brambilla}\ \emph {et~al.}(2005)\citenamefont
  {Brambilla}, \citenamefont {Pineda}, \citenamefont {Soto},\ and\
  \citenamefont {Vairo}}]{Brambilla:2004jw}%
  \BibitemOpen
  \bibfield  {author} {\bibinfo {author} {\bibfnamefont {N.}~\bibnamefont
  {Brambilla}}, \bibinfo {author} {\bibfnamefont {A.}~\bibnamefont {Pineda}},
  \bibinfo {author} {\bibfnamefont {J.}~\bibnamefont {Soto}}, \ and\ \bibinfo
  {author} {\bibfnamefont {A.}~\bibnamefont {Vairo}},\ }\href {\doibase
  10.1103/RevModPhys.77.1423} {\bibfield  {journal} {\bibinfo  {journal} {Rev.
  Mod. Phys.}\ }\textbf {\bibinfo {volume} {77}},\ \bibinfo {pages} {1423}
  (\bibinfo {year} {2005})},\ \Eprint {http://arxiv.org/abs/hep-ph/0410047}
  {arXiv:hep-ph/0410047 [hep-ph]} \BibitemShut {NoStop}%
\bibitem [{\citenamefont {Brambilla}\ \emph {et~al.}(2008)\citenamefont
  {Brambilla}, \citenamefont {Ghiglieri}, \citenamefont {Vairo},\ and\
  \citenamefont {Petreczky}}]{Brambilla:2008cx}%
  \BibitemOpen
  \bibfield  {author} {\bibinfo {author} {\bibfnamefont {N.}~\bibnamefont
  {Brambilla}}, \bibinfo {author} {\bibfnamefont {J.}~\bibnamefont
  {Ghiglieri}}, \bibinfo {author} {\bibfnamefont {A.}~\bibnamefont {Vairo}}, \
  and\ \bibinfo {author} {\bibfnamefont {P.}~\bibnamefont {Petreczky}},\ }\href
  {\doibase 10.1103/PhysRevD.78.014017} {\bibfield  {journal} {\bibinfo
  {journal} {Phys. Rev.}\ }\textbf {\bibinfo {volume} {D78}},\ \bibinfo {pages}
  {014017} (\bibinfo {year} {2008})},\ \Eprint {http://arxiv.org/abs/0804.0993}
  {0804.0993 [hep-ph]} \BibitemShut {NoStop}%
\bibitem [{\citenamefont {Laine}\ \emph {et~al.}(2007)\citenamefont {Laine},
  \citenamefont {Philipsen}, \citenamefont {Romatschke},\ and\ \citenamefont
  {Tassler}}]{Laine:2006ns}%
  \BibitemOpen
  \bibfield  {author} {\bibinfo {author} {\bibfnamefont {M.}~\bibnamefont
  {Laine}}, \bibinfo {author} {\bibfnamefont {O.}~\bibnamefont {Philipsen}},
  \bibinfo {author} {\bibfnamefont {P.}~\bibnamefont {Romatschke}}, \ and\
  \bibinfo {author} {\bibfnamefont {M.}~\bibnamefont {Tassler}},\ }\href
  {\doibase 10.1088/1126-6708/2007/03/054} {\bibfield  {journal} {\bibinfo
  {journal} {JHEP}\ }\textbf {\bibinfo {volume} {03}},\ \bibinfo {pages} {054}
  (\bibinfo {year} {2007})},\ \Eprint {http://arxiv.org/abs/hep-ph/0611300}
  {arXiv:hep-ph/0611300 [hep-ph]} \BibitemShut {NoStop}%
\bibitem [{\citenamefont {Rothkopf}\ \emph {et~al.}(2009)\citenamefont
  {Rothkopf}, \citenamefont {Hatsuda},\ and\ \citenamefont
  {Sasaki}}]{Rothkopf:2009pk}%
  \BibitemOpen
  \bibfield  {author} {\bibinfo {author} {\bibfnamefont {A.}~\bibnamefont
  {Rothkopf}}, \bibinfo {author} {\bibfnamefont {T.}~\bibnamefont {Hatsuda}}, \
  and\ \bibinfo {author} {\bibfnamefont {S.}~\bibnamefont {Sasaki}},\
  }\bibfield  {booktitle} {\emph {\bibinfo {booktitle} {{Proceedings, 27th
  International Symposium on Lattice field theory (Lattice 2009): Beijing, P.R.
  China, July 26-31, 2009}}},\ }\href@noop {} {\bibfield  {journal} {\bibinfo
  {journal} {PoS}\ }\textbf {\bibinfo {volume} {LAT2009}},\ \bibinfo {pages}
  {162} (\bibinfo {year} {2009})},\ \Eprint {http://arxiv.org/abs/0910.2321}
  {arXiv:0910.2321 [hep-lat]} \BibitemShut {NoStop}%
\bibitem [{\citenamefont {Rothkopf}\ \emph {et~al.}(2012)\citenamefont
  {Rothkopf}, \citenamefont {Hatsuda},\ and\ \citenamefont
  {Sasaki}}]{Rothkopf:2011db}%
  \BibitemOpen
  \bibfield  {author} {\bibinfo {author} {\bibfnamefont {A.}~\bibnamefont
  {Rothkopf}}, \bibinfo {author} {\bibfnamefont {T.}~\bibnamefont {Hatsuda}}, \
  and\ \bibinfo {author} {\bibfnamefont {S.}~\bibnamefont {Sasaki}},\ }\href
  {\doibase 10.1103/PhysRevLett.108.162001} {\bibfield  {journal} {\bibinfo
  {journal} {Phys. Rev. Lett.}\ }\textbf {\bibinfo {volume} {108}},\ \bibinfo
  {pages} {162001} (\bibinfo {year} {2012})},\ \Eprint
  {http://arxiv.org/abs/1108.1579} {arXiv:1108.1579 [hep-lat]} \BibitemShut
  {NoStop}%
\bibitem [{\citenamefont {Burnier}\ and\ \citenamefont
  {Rothkopf}(2012)}]{Burnier:2012az}%
  \BibitemOpen
  \bibfield  {author} {\bibinfo {author} {\bibfnamefont {Y.}~\bibnamefont
  {Burnier}}\ and\ \bibinfo {author} {\bibfnamefont {A.}~\bibnamefont
  {Rothkopf}},\ }\href {\doibase 10.1103/PhysRevD.86.051503} {\bibfield
  {journal} {\bibinfo  {journal} {Phys. Rev.}\ }\textbf {\bibinfo {volume}
  {D86}},\ \bibinfo {pages} {051503} (\bibinfo {year} {2012})},\ \Eprint
  {http://arxiv.org/abs/1208.1899} {arXiv:1208.1899 [hep-ph]} \BibitemShut
  {NoStop}%
\bibitem [{\citenamefont {Burnier}\ and\ \citenamefont
  {Rothkopf}(2013)}]{Burnier:2013nla}%
  \BibitemOpen
  \bibfield  {author} {\bibinfo {author} {\bibfnamefont {Y.}~\bibnamefont
  {Burnier}}\ and\ \bibinfo {author} {\bibfnamefont {A.}~\bibnamefont
  {Rothkopf}},\ }\href {\doibase 10.1103/PhysRevLett.111.182003} {\bibfield
  {journal} {\bibinfo  {journal} {Phys. Rev. Lett.}\ }\textbf {\bibinfo
  {volume} {111}},\ \bibinfo {pages} {182003} (\bibinfo {year} {2013})},\
  \Eprint {http://arxiv.org/abs/1307.6106} {arXiv:1307.6106 [hep-lat]}
  \BibitemShut {NoStop}%
\bibitem [{\citenamefont {Burnier}\ and\ \citenamefont
  {Rothkopf}(2017)}]{Burnier:2016mxc}%
  \BibitemOpen
  \bibfield  {author} {\bibinfo {author} {\bibfnamefont {Y.}~\bibnamefont
  {Burnier}}\ and\ \bibinfo {author} {\bibfnamefont {A.}~\bibnamefont
  {Rothkopf}},\ }\href {\doibase 10.1103/PhysRevD.95.054511} {\bibfield
  {journal} {\bibinfo  {journal} {Phys. Rev.}\ }\textbf {\bibinfo {volume}
  {D95}},\ \bibinfo {pages} {054511} (\bibinfo {year} {2017})},\ \Eprint
  {http://arxiv.org/abs/1607.04049} {arXiv:1607.04049 [hep-lat]} \BibitemShut
  {NoStop}%
\bibitem [{\citenamefont {Burnier}\ \emph {et~al.}(2015)\citenamefont
  {Burnier}, \citenamefont {Kaczmarek},\ and\ \citenamefont
  {Rothkopf}}]{Burnier:2015tda}%
  \BibitemOpen
  \bibfield  {author} {\bibinfo {author} {\bibfnamefont {Y.}~\bibnamefont
  {Burnier}}, \bibinfo {author} {\bibfnamefont {O.}~\bibnamefont {Kaczmarek}},
  \ and\ \bibinfo {author} {\bibfnamefont {A.}~\bibnamefont {Rothkopf}},\
  }\href {\doibase 10.1007/JHEP12(2015)101} {\bibfield  {journal} {\bibinfo
  {journal} {JHEP}\ }\textbf {\bibinfo {volume} {12}},\ \bibinfo {pages} {101}
  (\bibinfo {year} {2015})},\ \Eprint {http://arxiv.org/abs/1509.07366}
  {arXiv:1509.07366 [hep-ph]} \BibitemShut {NoStop}%
\bibitem [{\citenamefont {Burnier}\ \emph {et~al.}(2016)\citenamefont
  {Burnier}, \citenamefont {Kaczmarek},\ and\ \citenamefont
  {Rothkopf}}]{Burnier:2016kqm}%
  \BibitemOpen
  \bibfield  {author} {\bibinfo {author} {\bibfnamefont {Y.}~\bibnamefont
  {Burnier}}, \bibinfo {author} {\bibfnamefont {O.}~\bibnamefont {Kaczmarek}},
  \ and\ \bibinfo {author} {\bibfnamefont {A.}~\bibnamefont {Rothkopf}},\
  }\href {\doibase 10.1007/JHEP10(2016)032} {\bibfield  {journal} {\bibinfo
  {journal} {JHEP}\ }\textbf {\bibinfo {volume} {10}},\ \bibinfo {pages} {032}
  (\bibinfo {year} {2016})},\ \Eprint {http://arxiv.org/abs/1606.06211}
  {arXiv:1606.06211 [hep-ph]} \BibitemShut {NoStop}%
\bibitem [{\citenamefont {Burnier}\ and\ \citenamefont
  {Rothkopf}(2016)}]{Burnier:2015nsa}%
  \BibitemOpen
  \bibfield  {author} {\bibinfo {author} {\bibfnamefont {Y.}~\bibnamefont
  {Burnier}}\ and\ \bibinfo {author} {\bibfnamefont {A.}~\bibnamefont
  {Rothkopf}},\ }\href {\doibase 10.1016/j.physletb.2015.12.031} {\bibfield
  {journal} {\bibinfo  {journal} {Phys. Lett.}\ }\textbf {\bibinfo {volume}
  {B753}},\ \bibinfo {pages} {232} (\bibinfo {year} {2016})},\ \Eprint
  {http://arxiv.org/abs/1506.08684} {arXiv:1506.08684 [hep-ph]} \BibitemShut
  {NoStop}%
\bibitem [{\citenamefont {Romatschke}\ and\ \citenamefont
  {Strickland}(2004)}]{Romatschke:2004jh}%
  \BibitemOpen
  \bibfield  {author} {\bibinfo {author} {\bibfnamefont {P.}~\bibnamefont
  {Romatschke}}\ and\ \bibinfo {author} {\bibfnamefont {M.}~\bibnamefont
  {Strickland}},\ }\href {\doibase 10.1103/PhysRevD.70.116006} {\bibfield
  {journal} {\bibinfo  {journal} {Phys.Rev.}\ }\textbf {\bibinfo {volume}
  {D70}},\ \bibinfo {pages} {116006} (\bibinfo {year} {2004})},\ \Eprint
  {http://arxiv.org/abs/hep-ph/0406188} {arXiv:hep-ph/0406188 [hep-ph]}
  \BibitemShut {NoStop}%
\bibitem [{\citenamefont {Martinez}\ and\ \citenamefont
  {Strickland}(2011)}]{Martinez:2010sd}%
  \BibitemOpen
  \bibfield  {author} {\bibinfo {author} {\bibfnamefont {M.}~\bibnamefont
  {Martinez}}\ and\ \bibinfo {author} {\bibfnamefont {M.}~\bibnamefont
  {Strickland}},\ }\href {\doibase 10.1016/j.nuclphysa.2011.02.003} {\bibfield
  {journal} {\bibinfo  {journal} {Nucl.Phys.}\ }\textbf {\bibinfo {volume}
  {A856}},\ \bibinfo {pages} {68} (\bibinfo {year} {2011})},\ \Eprint
  {http://arxiv.org/abs/1011.3056} {arXiv:1011.3056 [nucl-th]} \BibitemShut
  {NoStop}%
\bibitem [{\citenamefont {Ryblewski}\ and\ \citenamefont
  {Florkowski}(2011{\natexlab{a}})}]{Ryblewski:2010bs}%
  \BibitemOpen
  \bibfield  {author} {\bibinfo {author} {\bibfnamefont {R.}~\bibnamefont
  {Ryblewski}}\ and\ \bibinfo {author} {\bibfnamefont {W.}~\bibnamefont
  {Florkowski}},\ }\href {\doibase 10.1088/0954-3899/38/1/015104} {\bibfield
  {journal} {\bibinfo  {journal} {J.Phys.G}\ }\textbf {\bibinfo {volume}
  {G38}},\ \bibinfo {pages} {015104} (\bibinfo {year} {2011}{\natexlab{a}})},\
  \Eprint {http://arxiv.org/abs/1007.4662} {arXiv:1007.4662 [nucl-th]}
  \BibitemShut {NoStop}%
\bibitem [{\citenamefont {Ryblewski}\ and\ \citenamefont
  {Florkowski}(2011{\natexlab{b}})}]{Ryblewski:2011aq}%
  \BibitemOpen
  \bibfield  {author} {\bibinfo {author} {\bibfnamefont {R.}~\bibnamefont
  {Ryblewski}}\ and\ \bibinfo {author} {\bibfnamefont {W.}~\bibnamefont
  {Florkowski}},\ }\href {\doibase 10.1140/epjc/s10052-011-1761-8} {\bibfield
  {journal} {\bibinfo  {journal} {Eur.Phys.J.}\ }\textbf {\bibinfo {volume}
  {C71}},\ \bibinfo {pages} {1761} (\bibinfo {year} {2011}{\natexlab{b}})},\
  \Eprint {http://arxiv.org/abs/1103.1260} {arXiv:1103.1260 [nucl-th]}
  \BibitemShut {NoStop}%
\bibitem [{\citenamefont {Martinez}\ \emph {et~al.}(2012)\citenamefont
  {Martinez}, \citenamefont {Ryblewski},\ and\ \citenamefont
  {Strickland}}]{Martinez:2012tu}%
  \BibitemOpen
  \bibfield  {author} {\bibinfo {author} {\bibfnamefont {M.}~\bibnamefont
  {Martinez}}, \bibinfo {author} {\bibfnamefont {R.}~\bibnamefont {Ryblewski}},
  \ and\ \bibinfo {author} {\bibfnamefont {M.}~\bibnamefont {Strickland}},\
  }\href {\doibase 10.1103/PhysRevC.85.064913} {\bibfield  {journal} {\bibinfo
  {journal} {Phys. Rev.}\ }\textbf {\bibinfo {volume} {C85}},\ \bibinfo {pages}
  {064913} (\bibinfo {year} {2012})},\ \Eprint {http://arxiv.org/abs/1204.1473}
  {arXiv:1204.1473 [nucl-th]} \BibitemShut {NoStop}%
\bibitem [{\citenamefont {Ryblewski}\ and\ \citenamefont
  {Florkowski}(2012)}]{Ryblewski:2012rr}%
  \BibitemOpen
  \bibfield  {author} {\bibinfo {author} {\bibfnamefont {R.}~\bibnamefont
  {Ryblewski}}\ and\ \bibinfo {author} {\bibfnamefont {W.}~\bibnamefont
  {Florkowski}},\ }\href {\doibase 10.1103/PhysRevC.85.064901} {\bibfield
  {journal} {\bibinfo  {journal} {Phys. Rev.}\ }\textbf {\bibinfo {volume}
  {C85}},\ \bibinfo {pages} {064901} (\bibinfo {year} {2012})},\ \Eprint
  {http://arxiv.org/abs/1204.2624} {arXiv:1204.2624 [nucl-th]} \BibitemShut
  {NoStop}%
\bibitem [{\citenamefont {Strickland}\ and\ \citenamefont
  {Yager-Elorriaga}(2010)}]{Strickland:2009ft}%
  \BibitemOpen
  \bibfield  {author} {\bibinfo {author} {\bibfnamefont {M.}~\bibnamefont
  {Strickland}}\ and\ \bibinfo {author} {\bibfnamefont {D.}~\bibnamefont
  {Yager-Elorriaga}},\ }\href {\doibase 10.1016/j.jcp.2010.04.032} {\bibfield
  {journal} {\bibinfo  {journal} {J.Comput.Phys.}\ }\textbf {\bibinfo {volume}
  {229}},\ \bibinfo {pages} {6015} (\bibinfo {year} {2010})},\ \Eprint
  {http://arxiv.org/abs/0904.0939} {arXiv:0904.0939 [quant-ph]} \BibitemShut
  {NoStop}%
\bibitem [{\citenamefont {Margotta}\ \emph {et~al.}(2011)\citenamefont
  {Margotta}, \citenamefont {McCarty}, \citenamefont {McGahan}, \citenamefont
  {Strickland},\ and\ \citenamefont {Yager-Elorriaga}}]{Margotta:2011ta}%
  \BibitemOpen
  \bibfield  {author} {\bibinfo {author} {\bibfnamefont {M.}~\bibnamefont
  {Margotta}}, \bibinfo {author} {\bibfnamefont {K.}~\bibnamefont {McCarty}},
  \bibinfo {author} {\bibfnamefont {C.}~\bibnamefont {McGahan}}, \bibinfo
  {author} {\bibfnamefont {M.}~\bibnamefont {Strickland}}, \ and\ \bibinfo
  {author} {\bibfnamefont {D.}~\bibnamefont {Yager-Elorriaga}},\ }\href
  {\doibase 10.1103/PhysRevD.83.105019, 10.1103/PhysRevD.84.069902,
  10.1103/PhysRevD.83.105019, 10.1103/PhysRevD.84.069902} {\bibfield  {journal}
  {\bibinfo  {journal} {Phys.Rev.}\ }\textbf {\bibinfo {volume} {D83}},\
  \bibinfo {pages} {105019} (\bibinfo {year} {2011})},\ \Eprint
  {http://arxiv.org/abs/1101.4651} {arXiv:1101.4651 [hep-ph]} \BibitemShut
  {NoStop}%
\bibitem [{\citenamefont {Florkowski}\ \emph
  {et~al.}(2013{\natexlab{a}})\citenamefont {Florkowski}, \citenamefont
  {Ryblewski},\ and\ \citenamefont {Strickland}}]{Florkowski:2013lza}%
  \BibitemOpen
  \bibfield  {author} {\bibinfo {author} {\bibfnamefont {W.}~\bibnamefont
  {Florkowski}}, \bibinfo {author} {\bibfnamefont {R.}~\bibnamefont
  {Ryblewski}}, \ and\ \bibinfo {author} {\bibfnamefont {M.}~\bibnamefont
  {Strickland}},\ }\href {\doibase 10.1016/j.nuclphysa.2013.08.004} {\bibfield
  {journal} {\bibinfo  {journal} {Nucl.Phys.}\ }\textbf {\bibinfo {volume}
  {A916}},\ \bibinfo {pages} {249} (\bibinfo {year} {2013}{\natexlab{a}})},\
  \Eprint {http://arxiv.org/abs/1304.0665} {arXiv:1304.0665 [nucl-th]}
  \BibitemShut {NoStop}%
\bibitem [{\citenamefont {Florkowski}\ \emph
  {et~al.}(2013{\natexlab{b}})\citenamefont {Florkowski}, \citenamefont
  {Ryblewski},\ and\ \citenamefont {Strickland}}]{Florkowski:2013lya}%
  \BibitemOpen
  \bibfield  {author} {\bibinfo {author} {\bibfnamefont {W.}~\bibnamefont
  {Florkowski}}, \bibinfo {author} {\bibfnamefont {R.}~\bibnamefont
  {Ryblewski}}, \ and\ \bibinfo {author} {\bibfnamefont {M.}~\bibnamefont
  {Strickland}},\ }\href {\doibase 10.1103/PhysRevC.88.024903} {\bibfield
  {journal} {\bibinfo  {journal} {Phys. Rev.}\ }\textbf {\bibinfo {volume}
  {C88}},\ \bibinfo {pages} {024903} (\bibinfo {year} {2013}{\natexlab{b}})},\
  \Eprint {http://arxiv.org/abs/1305.7234} {arXiv:1305.7234 [nucl-th]}
  \BibitemShut {NoStop}%
\bibitem [{\citenamefont {Bazow}\ \emph {et~al.}(2014)\citenamefont {Bazow},
  \citenamefont {Heinz},\ and\ \citenamefont {Strickland}}]{Bazow:2013ifa}%
  \BibitemOpen
  \bibfield  {author} {\bibinfo {author} {\bibfnamefont {D.}~\bibnamefont
  {Bazow}}, \bibinfo {author} {\bibfnamefont {U.~W.}\ \bibnamefont {Heinz}}, \
  and\ \bibinfo {author} {\bibfnamefont {M.}~\bibnamefont {Strickland}},\
  }\href {\doibase 10.1103/PhysRevC.90.054910} {\bibfield  {journal} {\bibinfo
  {journal} {Phys. Rev.}\ }\textbf {\bibinfo {volume} {C90}},\ \bibinfo {pages}
  {054910} (\bibinfo {year} {2014})},\ \Eprint {http://arxiv.org/abs/1311.6720}
  {arXiv:1311.6720 [nucl-th]} \BibitemShut {NoStop}%
\bibitem [{\citenamefont {Florkowski}\ \emph {et~al.}(2014)\citenamefont
  {Florkowski}, \citenamefont {Maksymiuk}, \citenamefont {Ryblewski},\ and\
  \citenamefont {Strickland}}]{Florkowski:2014sfa}%
  \BibitemOpen
  \bibfield  {author} {\bibinfo {author} {\bibfnamefont {W.}~\bibnamefont
  {Florkowski}}, \bibinfo {author} {\bibfnamefont {E.}~\bibnamefont
  {Maksymiuk}}, \bibinfo {author} {\bibfnamefont {R.}~\bibnamefont
  {Ryblewski}}, \ and\ \bibinfo {author} {\bibfnamefont {M.}~\bibnamefont
  {Strickland}},\ }\href {\doibase 10.1103/PhysRevC.89.054908} {\bibfield
  {journal} {\bibinfo  {journal} {Phys.Rev.}\ }\textbf {\bibinfo {volume}
  {C89}},\ \bibinfo {pages} {054908} (\bibinfo {year} {2014})},\ \Eprint
  {http://arxiv.org/abs/1402.7348} {arXiv:1402.7348 [hep-ph]} \BibitemShut
  {NoStop}%
\bibitem [{\citenamefont {Florkowski}\ and\ \citenamefont
  {Maksymiuk}(2014)}]{Florkowski:2014sda}%
  \BibitemOpen
  \bibfield  {author} {\bibinfo {author} {\bibfnamefont {W.}~\bibnamefont
  {Florkowski}}\ and\ \bibinfo {author} {\bibfnamefont {E.}~\bibnamefont
  {Maksymiuk}},\ }\href@noop {} {\  (\bibinfo {year} {2014})},\ \Eprint
  {http://arxiv.org/abs/1411.3666} {arXiv:1411.3666 [hep-ph]} \BibitemShut
  {NoStop}%
\bibitem [{\citenamefont {Denicol}\ \emph
  {et~al.}(2014{\natexlab{a}})\citenamefont {Denicol}, \citenamefont {Heinz},
  \citenamefont {Martinez}, \citenamefont {Noronha},\ and\ \citenamefont
  {Strickland}}]{Denicol:2014xca}%
  \BibitemOpen
  \bibfield  {author} {\bibinfo {author} {\bibfnamefont {G.~S.}\ \bibnamefont
  {Denicol}}, \bibinfo {author} {\bibfnamefont {U.~W.}\ \bibnamefont {Heinz}},
  \bibinfo {author} {\bibfnamefont {M.}~\bibnamefont {Martinez}}, \bibinfo
  {author} {\bibfnamefont {J.}~\bibnamefont {Noronha}}, \ and\ \bibinfo
  {author} {\bibfnamefont {M.}~\bibnamefont {Strickland}},\ }\href {\doibase
  10.1103/PhysRevLett.113.202301} {\bibfield  {journal} {\bibinfo  {journal}
  {Phys.Rev.Lett.}\ }\textbf {\bibinfo {volume} {113}},\ \bibinfo {pages}
  {202301} (\bibinfo {year} {2014}{\natexlab{a}})},\ \Eprint
  {http://arxiv.org/abs/1408.5646} {arXiv:1408.5646 [hep-ph]} \BibitemShut
  {NoStop}%
\bibitem [{\citenamefont {Denicol}\ \emph
  {et~al.}(2014{\natexlab{b}})\citenamefont {Denicol}, \citenamefont {Heinz},
  \citenamefont {Martinez}, \citenamefont {Noronha},\ and\ \citenamefont
  {Strickland}}]{Denicol:2014tha}%
  \BibitemOpen
  \bibfield  {author} {\bibinfo {author} {\bibfnamefont {G.~S.}\ \bibnamefont
  {Denicol}}, \bibinfo {author} {\bibfnamefont {U.~W.}\ \bibnamefont {Heinz}},
  \bibinfo {author} {\bibfnamefont {M.}~\bibnamefont {Martinez}}, \bibinfo
  {author} {\bibfnamefont {J.}~\bibnamefont {Noronha}}, \ and\ \bibinfo
  {author} {\bibfnamefont {M.}~\bibnamefont {Strickland}},\ }\href {\doibase
  10.1103/PhysRevD.90.125026} {\bibfield  {journal} {\bibinfo  {journal}
  {Phys.Rev.}\ }\textbf {\bibinfo {volume} {D90}},\ \bibinfo {pages} {125026}
  (\bibinfo {year} {2014}{\natexlab{b}})},\ \Eprint
  {http://arxiv.org/abs/1408.7048} {arXiv:1408.7048 [hep-ph]} \BibitemShut
  {NoStop}%
\bibitem [{\citenamefont {Nopoush}\ \emph
  {et~al.}(2014{\natexlab{b}})\citenamefont {Nopoush}, \citenamefont
  {Ryblewski},\ and\ \citenamefont {Strickland}}]{Nopoush:2014qba}%
  \BibitemOpen
  \bibfield  {author} {\bibinfo {author} {\bibfnamefont {M.}~\bibnamefont
  {Nopoush}}, \bibinfo {author} {\bibfnamefont {R.}~\bibnamefont {Ryblewski}},
  \ and\ \bibinfo {author} {\bibfnamefont {M.}~\bibnamefont {Strickland}},\
  }\href@noop {} {\  (\bibinfo {year} {2014}{\natexlab{b}})},\ \Eprint
  {http://arxiv.org/abs/1410.6790} {arXiv:1410.6790 [nucl-th]} \BibitemShut
  {NoStop}%
\bibitem [{\citenamefont {Molnar}\ \emph {et~al.}(2016)\citenamefont {Molnar},
  \citenamefont {Niemi},\ and\ \citenamefont {Rischke}}]{Molnar:2016gwq}%
  \BibitemOpen
  \bibfield  {author} {\bibinfo {author} {\bibfnamefont {E.}~\bibnamefont
  {Molnar}}, \bibinfo {author} {\bibfnamefont {H.}~\bibnamefont {Niemi}}, \
  and\ \bibinfo {author} {\bibfnamefont {D.~H.}\ \bibnamefont {Rischke}},\
  }\href {\doibase 10.1103/PhysRevD.94.125003} {\bibfield  {journal} {\bibinfo
  {journal} {Phys. Rev.}\ }\textbf {\bibinfo {volume} {D94}},\ \bibinfo {pages}
  {125003} (\bibinfo {year} {2016})},\ \Eprint
  {http://arxiv.org/abs/1606.09019} {arXiv:1606.09019 [nucl-th]} \BibitemShut
  {NoStop}%
\bibitem [{\citenamefont {Martinez}\ \emph {et~al.}(2017)\citenamefont
  {Martinez}, \citenamefont {McNelis},\ and\ \citenamefont
  {Heinz}}]{Martinez:2017ibh}%
  \BibitemOpen
  \bibfield  {author} {\bibinfo {author} {\bibfnamefont {M.}~\bibnamefont
  {Martinez}}, \bibinfo {author} {\bibfnamefont {M.}~\bibnamefont {McNelis}}, \
  and\ \bibinfo {author} {\bibfnamefont {U.}~\bibnamefont {Heinz}},\ }\href
  {\doibase 10.1103/PhysRevC.95.054907} {\bibfield  {journal} {\bibinfo
  {journal} {Phys. Rev.}\ }\textbf {\bibinfo {volume} {C95}},\ \bibinfo {pages}
  {054907} (\bibinfo {year} {2017})},\ \Eprint
  {http://arxiv.org/abs/1703.10955} {arXiv:1703.10955 [nucl-th]} \BibitemShut
  {NoStop}%
\bibitem [{\citenamefont {Damodaran}\ \emph {et~al.}(2017)\citenamefont
  {Damodaran}, \citenamefont {Molnar}, \citenamefont {Barnaföldi},
  \citenamefont {Berényi},\ and\ \citenamefont {Ferenc
  Nagy-Egri}}]{Damodaran:2017ior}%
  \BibitemOpen
  \bibfield  {author} {\bibinfo {author} {\bibfnamefont {M.}~\bibnamefont
  {Damodaran}}, \bibinfo {author} {\bibfnamefont {D.}~\bibnamefont {Molnar}},
  \bibinfo {author} {\bibfnamefont {G.~G.}\ \bibnamefont {Barnaföldi}},
  \bibinfo {author} {\bibfnamefont {D.}~\bibnamefont {Berényi}}, \ and\
  \bibinfo {author} {\bibfnamefont {M.}~\bibnamefont {Ferenc Nagy-Egri}},\
  }\href@noop {} {\  (\bibinfo {year} {2017})},\ \Eprint
  {http://arxiv.org/abs/1707.00793} {arXiv:1707.00793 [nucl-th]} \BibitemShut
  {NoStop}%
\bibitem [{\citenamefont {Ryblewski}\ and\ \citenamefont
  {Strickland}(2015)}]{Ryblewski:2015hea}%
  \BibitemOpen
  \bibfield  {author} {\bibinfo {author} {\bibfnamefont {R.}~\bibnamefont
  {Ryblewski}}\ and\ \bibinfo {author} {\bibfnamefont {M.}~\bibnamefont
  {Strickland}},\ }\href {\doibase 10.1103/PhysRevD.92.025026} {\bibfield
  {journal} {\bibinfo  {journal} {Phys. Rev.}\ }\textbf {\bibinfo {volume}
  {D92}},\ \bibinfo {pages} {025026} (\bibinfo {year} {2015})},\ \Eprint
  {http://arxiv.org/abs/1501.03418} {arXiv:1501.03418 [nucl-th]} \BibitemShut
  {NoStop}%
\bibitem [{\citenamefont {Bozek}\ and\ \citenamefont
  {Wyskiel}(2010)}]{Bozek:2010bi}%
  \BibitemOpen
  \bibfield  {author} {\bibinfo {author} {\bibfnamefont {P.}~\bibnamefont
  {Bozek}}\ and\ \bibinfo {author} {\bibfnamefont {I.}~\bibnamefont
  {Wyskiel}},\ }\href {\doibase 10.1103/PhysRevC.81.054902} {\bibfield
  {journal} {\bibinfo  {journal} {Phys. Rev.}\ }\textbf {\bibinfo {volume}
  {C81}},\ \bibinfo {pages} {054902} (\bibinfo {year} {2010})},\ \Eprint
  {http://arxiv.org/abs/1002.4999} {arXiv:1002.4999 [nucl-th]} \BibitemShut
  {NoStop}%
\bibitem [{\citenamefont {Karsch}\ and\ \citenamefont
  {Petronzio}(1987)}]{Karsch:1987uk}%
  \BibitemOpen
  \bibfield  {author} {\bibinfo {author} {\bibfnamefont {F.}~\bibnamefont
  {Karsch}}\ and\ \bibinfo {author} {\bibfnamefont {R.}~\bibnamefont
  {Petronzio}},\ }\href {\doibase 10.1016/0370-2693(87)90465-5} {\bibfield
  {journal} {\bibinfo  {journal} {Phys. Lett.}\ }\textbf {\bibinfo {volume}
  {B193}},\ \bibinfo {pages} {105} (\bibinfo {year} {1987})}\BibitemShut
  {NoStop}%
\bibitem [{\citenamefont {W\"{o}eri}(2014)}]{Woeri:2015hq}%
  \BibitemOpen
  \bibfield  {author} {\bibinfo {author} {\bibfnamefont {H.}~\bibnamefont
  {W\"{o}eri}},\ }\href@noop {} {\enquote {\bibinfo {title} {{International
  Workshop on Heavy Quarkonium 2014}},}\ }\bibinfo {howpublished}
  {\url{https://indico.cern.ch/event/278195/session/7/contribution/104/material/slides/0.pdf}}
  (\bibinfo {year} {2014})\BibitemShut {NoStop}%
\bibitem [{\citenamefont {Rothkopf}(2014)}]{Rothkopf:2013kya}%
  \BibitemOpen
  \bibfield  {author} {\bibinfo {author} {\bibfnamefont {A.}~\bibnamefont
  {Rothkopf}},\ }\href {\doibase 10.1007/JHEP04(2014)085} {\bibfield  {journal}
  {\bibinfo  {journal} {JHEP}\ }\textbf {\bibinfo {volume} {04}},\ \bibinfo
  {pages} {085} (\bibinfo {year} {2014})},\ \Eprint
  {http://arxiv.org/abs/1312.3246} {arXiv:1312.3246 [hep-ph]} \BibitemShut
  {NoStop}%
\bibitem [{\citenamefont {Kajimoto}\ \emph {et~al.}(2017)\citenamefont
  {Kajimoto}, \citenamefont {Akamatsu}, \citenamefont {Asakawa},\ and\
  \citenamefont {Rothkopf}}]{Kajimoto:2017rel}%
  \BibitemOpen
  \bibfield  {author} {\bibinfo {author} {\bibfnamefont {S.}~\bibnamefont
  {Kajimoto}}, \bibinfo {author} {\bibfnamefont {Y.}~\bibnamefont {Akamatsu}},
  \bibinfo {author} {\bibfnamefont {M.}~\bibnamefont {Asakawa}}, \ and\
  \bibinfo {author} {\bibfnamefont {A.}~\bibnamefont {Rothkopf}},\ }\href@noop
  {} {\  (\bibinfo {year} {2017})},\ \Eprint {http://arxiv.org/abs/1705.03365}
  {arXiv:1705.03365 [nucl-th]} \BibitemShut {NoStop}%
\bibitem [{\citenamefont {Akamatsu}\ and\ \citenamefont
  {Rothkopf}(2012)}]{Akamatsu:2011se}%
  \BibitemOpen
  \bibfield  {author} {\bibinfo {author} {\bibfnamefont {Y.}~\bibnamefont
  {Akamatsu}}\ and\ \bibinfo {author} {\bibfnamefont {A.}~\bibnamefont
  {Rothkopf}},\ }\href {\doibase 10.1103/PhysRevD.85.105011} {\bibfield
  {journal} {\bibinfo  {journal} {Phys. Rev.}\ }\textbf {\bibinfo {volume}
  {D85}},\ \bibinfo {pages} {105011} (\bibinfo {year} {2012})},\ \Eprint
  {http://arxiv.org/abs/1110.1203} {arXiv:1110.1203 [hep-ph]} \BibitemShut
  {NoStop}%
\bibitem [{\citenamefont {Akamatsu}(2015{\natexlab{a}})}]{Akamatsu:2014qsa}%
  \BibitemOpen
  \bibfield  {author} {\bibinfo {author} {\bibfnamefont {Y.}~\bibnamefont
  {Akamatsu}},\ }\href {\doibase 10.1103/PhysRevD.91.056002} {\bibfield
  {journal} {\bibinfo  {journal} {Phys. Rev.}\ }\textbf {\bibinfo {volume}
  {D91}},\ \bibinfo {pages} {056002} (\bibinfo {year} {2015}{\natexlab{a}})},\
  \Eprint {http://arxiv.org/abs/1403.5783} {arXiv:1403.5783 [hep-ph]}
  \BibitemShut {NoStop}%
\bibitem [{\citenamefont {Akamatsu}(2015{\natexlab{b}})}]{Akamatsu:2015kaa}%
  \BibitemOpen
  \bibfield  {author} {\bibinfo {author} {\bibfnamefont {Y.}~\bibnamefont
  {Akamatsu}},\ }\href {\doibase 10.1103/PhysRevC.92.044911} {\bibfield
  {journal} {\bibinfo  {journal} {Phys. Rev.}\ }\textbf {\bibinfo {volume}
  {C92}},\ \bibinfo {pages} {044911} (\bibinfo {year} {2015}{\natexlab{b}})},\
  \Eprint {http://arxiv.org/abs/1503.08110} {arXiv:1503.08110 [nucl-th]}
  \BibitemShut {NoStop}%
\bibitem [{\citenamefont {De~Boni}(2017)}]{DeBoni:2017ocl}%
  \BibitemOpen
  \bibfield  {author} {\bibinfo {author} {\bibfnamefont {D.}~\bibnamefont
  {De~Boni}},\ }\href {\doibase 10.1007/JHEP08(2017)064} {\bibfield  {journal}
  {\bibinfo  {journal} {JHEP}\ }\textbf {\bibinfo {volume} {08}},\ \bibinfo
  {pages} {064} (\bibinfo {year} {2017})},\ \Eprint
  {http://arxiv.org/abs/1705.03567} {arXiv:1705.03567 [hep-ph]} \BibitemShut
  {NoStop}%
\bibitem [{\citenamefont {Brambilla}\ \emph {et~al.}(2016)\citenamefont
  {Brambilla}, \citenamefont {Escobedo}, \citenamefont {Soto},\ and\
  \citenamefont {Vairo}}]{Brambilla:2016wgg}%
  \BibitemOpen
  \bibfield  {author} {\bibinfo {author} {\bibfnamefont {N.}~\bibnamefont
  {Brambilla}}, \bibinfo {author} {\bibfnamefont {M.~A.}\ \bibnamefont
  {Escobedo}}, \bibinfo {author} {\bibfnamefont {J.}~\bibnamefont {Soto}}, \
  and\ \bibinfo {author} {\bibfnamefont {A.}~\bibnamefont {Vairo}},\
  }\href@noop {} {\  (\bibinfo {year} {2016})},\ \Eprint
  {http://arxiv.org/abs/1612.07248} {arXiv:1612.07248 [hep-ph]} \BibitemShut
  {NoStop}%
\end{thebibliography}%

\end{document}